\documentclass[twocolumn]{aastex631}

\begin{document}

\title{The BANANA Project.\ VII.\ High Eccentricity Predicts Spin-Orbit Misalignment in Binaries}

\author[0000-0003-2173-0689]{Marcus L.\ Marcussen}
\affil{Department of Physics and Astronomy, Aarhus University, Ny Munkegade 120, 8000 Aarhus C, Denmark}
\affil{Department of Astrophysical Sciences, Peyton Hall, 4 Ivy Lane, Princeton, NJ 08544, USA}

\email{marcuslmarcussen@gmail.com}

\author[0000-0003-1762-8235]{Simon H.\ Albrecht}
\affil{Department of Physics and Astronomy, Aarhus University, Ny Munkegade 120, 8000 Aarhus C, Denmark}
\affil{Department of Astrophysical Sciences, Peyton Hall, 4 Ivy Lane, Princeton, NJ 08544, USA}

\author[0000-0002-4265-047X]{Joshua N.\ Winn}
\affil{Department of Astrophysical Sciences, Peyton Hall, 4 Ivy Lane, Princeton, NJ 08544, USA}

\author[0000-0001-8283-3425]{Yubo Su}
\affil{Department of Astrophysical Sciences, Peyton Hall, 4 Ivy Lane, Princeton, NJ 08544, USA}

\author[0000-0002-8661-2571]{Mia S.\ Lundkvist}
\affil{Department of Physics and Astronomy, Aarhus University, Ny Munkegade 120, 8000 Aarhus C, Denmark}

\author[0000-0001-5761-6779]{Kevin C.\ Schlaufman}
\affiliation{William H.\ Miller III Department of Physics \& Astronomy,
Johns Hopkins University, 3400 N Charles St, Baltimore, MD 21218, USA}

\begin{abstract}

The degree of spin-orbit alignment in a population of binary stars can be determined from measurements of their orbital inclinations and rotational broadening of their spectral lines.
Alignment in a face-on binary guarantees low rotational broadening, while alignment in an edge-on binary maximizes the rotational broadening.
In contrast, if spin-orbit angles ($\psi$) are random,
rotational broadening should not depend on orbital inclination.
Using this technique, we investigated a sample of 2{,}727 astrometric binaries from Gaia DR3
with F-type primaries and orbital periods between 50 and 1000 days (separations 0.3--2.7~au).
We found that $\psi$ is strongly
associated with $e$, the orbital eccentricity.
When $e<0.15$, the mean spin-orbit angle is $\langle\psi\rangle = 6.9_{-4.1}^{+5.4}$\,degrees, while for $e>0.7$, it rises to
$\langle\psi\rangle = 46_{-24}^{+26}$\,degrees. These results suggest that some binaries are affected by processes during their formation or evolution that excite both orbital eccentricity and inclination.
\end{abstract}

\section{Introduction} 

Stellar spin-orbit angles (also called obliquities) are affected by processes that take place during the formation and evolution of binary stars. Binaries that formed via disk fragmentation are expected to have well-aligned rotational and orbital axes. However, numerous processes can misalign a binary, such as chaotic accretion during the star formation process \citep{bate2010,Thies+2011,fielding2015,Offner+2016,bate2018,JenningsChiang2021},
torques from a warped circumbinary disk \citep{AndersonLai2021}, close encounters with other stars in the birth cluster \citep[e.g.][]{1996HeggieRasio,2021rodetsulai},
and Kozai-Lidov cycles caused by a distant third star \citep[e.g.\ ][]{mazeh1979,eggleton2001,fabrycky2007,naoz2014,AndersonLaiStorch2017}.

However, there are relatively few observational constraints on spin-orbit alignment in binaries. Most existing measurements of spin-orbit angles are not for the spins of stars with respect to binary orbital planes, but rather for
the spins of stars relative to the plane of a \textit{planetary orbit} \citep{AlbrechtDawsonWinn2022}.
The aim of project BANANA (Binaries Are Not Always Neatly Aligned) is to measure the obliquities of stars in binary systems and thereby constrain theories of binary formation and evolution.  For the close double star systems DI\,Herculis and CV\,Vel, we found that all four stars have large obliquities \citep{albrecht2009,Albrecht2014CVVel}.
We have also found some systems to
be closely aligned \citep{albrecht2007,Albrecht+2011,albrecht2013,albrecht2014,MarcussenAlbrecht2022}. We refer the reader to \cite{pavlovski2011,triaud2013,lehman2013,Philippov+2013,zhou2013,sybilski2018} and \cite{Ball+2023} for descriptions of similar efforts by other groups to measure spin-orbit angles in close binaries.  

Spin-orbit alignment has also been studied in wider binaries, with separations exceeding 1~au \citep[see, e.g.,][]{weis1974,hale1994,glebocki1997,howe2009,JustesenAlbrecht2020}. However, \cite{JustesenAlbrecht2019} showed that with the data at hand, it is still too early to draw firm conclusions about
spin-orbit angles in wide binaries. See \cite{Offner+2023} for a recent review on multiple star formation and measurements of the angles between protostars and their disks.  

Here we present a study of spin-orbit alignment in a population of several thousand wide binaries that were
selected from Data Release 3 (DR3) of the Gaia mission \citep{GaiaDR3}. 
For reasons relating to our technique,
described below, the sample is
restricted to single-lined binaries with F-type primaries and orbital periods ranging from $50$ to $1000$ days. 
Our technique builds on previous work by
many others \cite[e.g.\ ][]{weis1974,hale1994,schlaufman2010,MasudaWinn2020,Louden+2021}, in that we rely on observations of rotational broadening of spectral lines.
The $\sin i$ dependence of rotational
broadening can be used to extract statistical information about
the orientation distribution of a population of stars. We combine this
information with 
orbital inclinations derived from
astrometric data to arrive at constraints
on the distribution of spin-orbit angles.

Section~\ref{sec:sample} describes our selection of Gaia DR3 binaries.
Section~\ref{sec:vbroad_diff} compares the spectral line broadening observed for face-on binaries and edge-on binaries.
A significant difference was found --- implying low obliquities ---
but only for binaries with relatively
low orbital eccentricities ($e\lesssim 0.5$).
Section \ref{sec:method} describes our ``forward-modeling'' method for deriving quantitative constraints on the obliquity distribution. Section \ref{sec:results} displays the results of this method, 
and Section~\ref{sec:discussion} discusses possible implications for theories of binary formation and dynamics. We also highlight future directions of research using this technique.

\section{Sample} 
\label{sec:sample}

\begin{figure}
  \begin{center}
    \includegraphics[width=1\columnwidth]{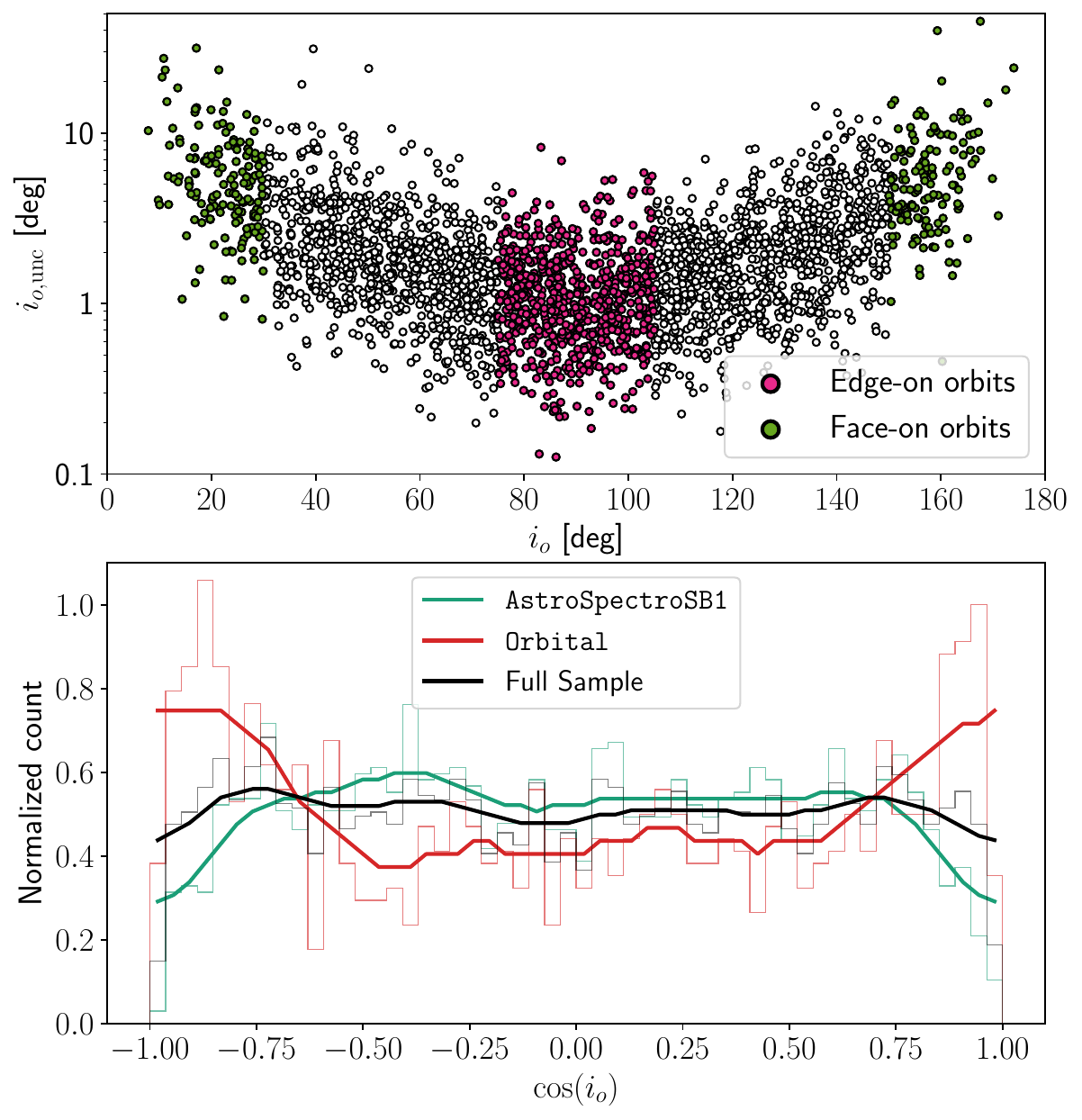}
    \caption{
    \label{fig:inclination_distribution} 
    \textbf{Orbital inclinations $i_o$ of binaries in the sample.}
    {\it Top:} Values and uncertainties of $i_o$ from Gaia DR3.
    We classified the binaries as ``edge-on'' (pink points), ``face-on''
    (green points) or intermediate (white points). {\it Bottom:} Distribution of $\cos{i_o}$ of stars for which 
    SB1 solutions are available in addition to astrometry (\texttt{AstroSpectroSB1})
    and stars for which only astrometric data are available (\texttt{Orbital}). An isotropic distribution of orientations would lead to a uniform distribution in $\cos i_o$.
    }
  \end{center}
\end{figure}

This work is based on the catalog of astrometric binaries that accompanied
Gaia DR3 \citep{Gaia2022Teaser}. For all of the binaries, the Keplerian orbital elements were determined by fitting the time-series astrometric data. The astrometry-only solutions are labelled \texttt{Orbital} in the Non-Single Source Gaia DR3 table \citep{Gaia2022Teaser}. All of the binaries were also observed
with Gaia's Radial Velocity Spectrometer (RVS) and a subset were found to be single-lined binaries; for those systems, data are available
from the joint astrometric and spectroscopic solutions, \texttt{AstroSpectroSB1}.

For our study, a key parameter from Gaia DR3 is the \texttt{vbroad} parameter \citep{GaiaVbroad2023}, which quantifies
the observed spectral-line broadening of the primary star.
As we will show below, for most
F-type stars, the broadening is
dominated by rotational
line broadening, allowing for
the calibration of a relationship
between \texttt{vbroad} and
the projected rotation velocity
$v\sin i$, where $i$ is the inclination
of the stellar rotation axis. The sample we study includes only single-lined binaries,
for which the observed flux comes
mainly from the primary star. The tabulated \texttt{vbroad} parameter
can therefore safely be attributed to
the primary star.
Thus, in what follows, when we refer to spin-orbit alignment, we mean the alignment between the spin axis of the \textit{primary} star and the axis of the binary orbit.

\begin{figure*}
    \centering
    \includegraphics[width = 1\textwidth]{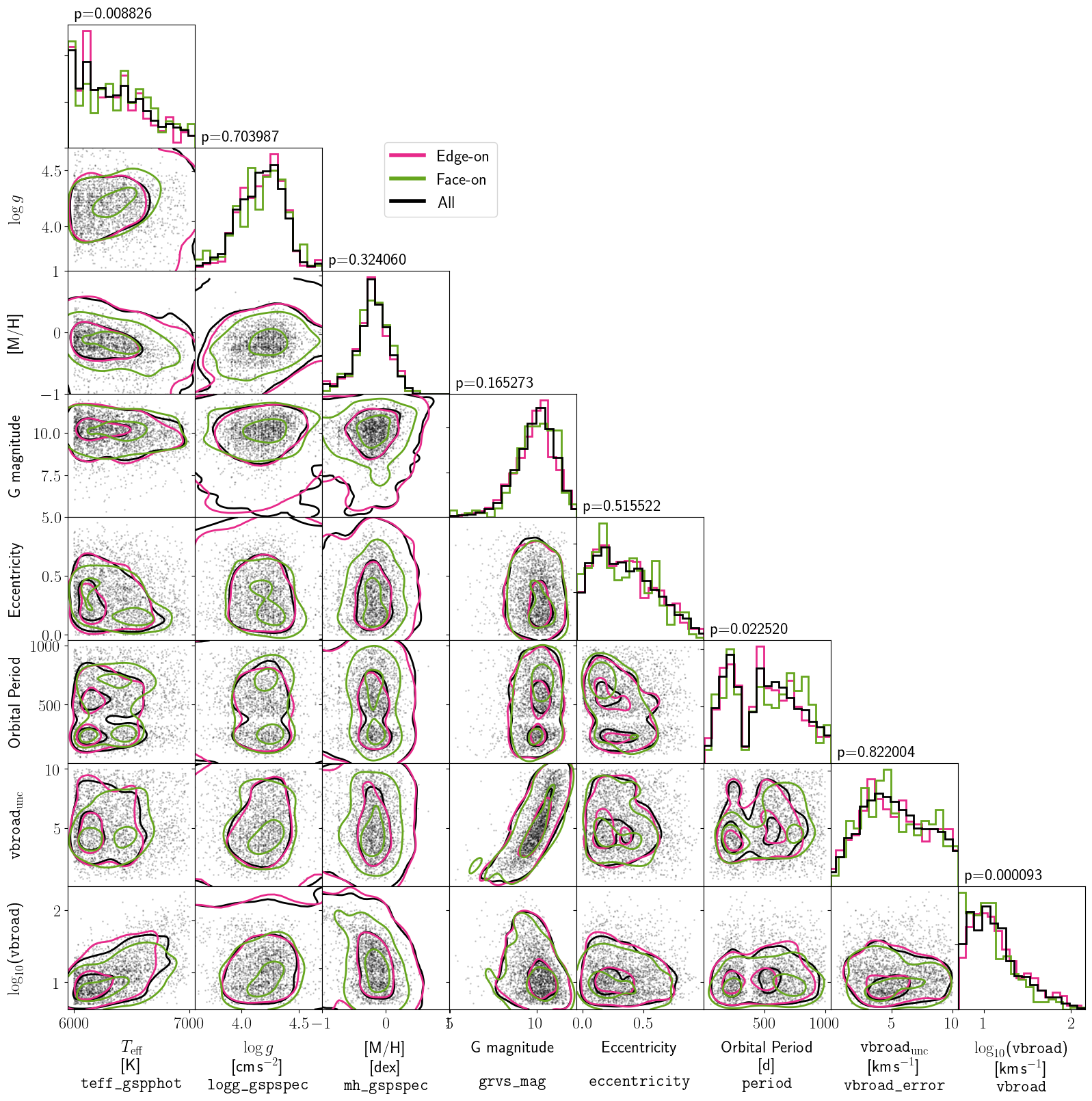}
    \caption{\textbf{Physical parameters of the binary sample.} A selection of relevant parameters and their correlations are shown for the full sample of binary systems (black) and for the face-on and edge-on subsets (green and pink). The $p$-values printed above the histograms come from KS tests comparing the edge-on and face-on samples.}
    \label{fig:sample_parameters}
\end{figure*} 

We imposed several selection criteria to create a relatively homogeneous and
high-quality sample.
We started with systems for which the ${\rm G}_{\rm RVS}$ magnitude is brighter than 12 because the $\texttt{vbroad}$ parameter is not available for fainter stars.
We required the reported uncertainty
in \texttt{vbroad} to be less than $10$\,km\,s$^{-1}$.
We restricted the range of orbital
periods to be between 50 and 1000~days. We omitted stars with unusually low or high surface gravity or metallicity. Specifically, we restricted $\log{g}$ to the range between 3.8 and 4.5 and metallicity $[$M$/$H$]$ to the range between $-1$ and 1.
We required the effective temperature $T_\mathrm{eff}$ to be between
6000\,K and 7000\,K, roughly corresponding to F-type stars, and
we also required the uncertainty to
be smaller than 100\,K. Main sequence stars with lower effective temperatures tend to rotate too slowly
to allow for reliable observations of rotational broadening
with the RVS.
For stars hotter than $\sim\,7000\,$K, \cite{GaiaVbroad2023}
found the \texttt{vbroad} parameter to
be an unreliable measure of rotational broadening.
Finally, we required the ``significance'' parameter $s$ associated with the astrometric solution to exceed 5. Roughly speaking, the $s$ parameter is the signal-to-noise ratio \citep{Halbwachs+2023}. A total of 2{,}727 systems satisfied all of our criteria, which are summarized as:
\begin{align*}
&\ \texttt{vbroad}_{\rm unc} <10\,\mathrm{km}\,\mathrm{s}^{-1}\\
50\,\mathrm{d} < &\ \text{orbital period} < 1000\,\mathrm{d}\\
3.8 < &\ \log g < 4.5\\
-1 < &\ [\text{M/H}] < 1\\
6000\,\rm{K} < &\ T_{\text{eff}} < 7000\,\text{K} \\
&\ T_{\rm eff, unc} <100\,\mathrm{K}\\
5< &\ \text{significance $s$ of astrometric solution}\\
\end{align*}

The top panel of Figure~\ref{fig:inclination_distribution} shows the orbital inclinations of this sample, which we obtained by transforming the tabulated Thiele-Innes elements into Campbell elements \citep{Gaia2022Teaser}. Not surprisingly, the uncertainties tend to be larger when the orbit is nearly face-on. The bottom panel shows the distributions of $\cos i_{\rm o}$ for the \texttt{AstroSpectroSB1} binaries and for the \texttt{Orbital} binaries. The \texttt{AstroSpectroSB1} contains relatively few face-on orbits, as expected, since a low inclination implies low radial velocities. The distribution of $\cos i_{\rm o}$ is close to uniform for all the binaries (i.e.\ the joint sample of \texttt{Orbital} and \texttt{AstroSpectroSB1}), corresponding to an isotropic distribution of orientations in three dimensions. 

We classified each of the 2{,}727 systems based on orbital inclination. ``Edge-on'' systems have inclinations
between $75^\circ$ and $105^\circ$.
``Face-on'' systems have inclinations
that are either between $0^\circ$ and $30^\circ$ or between $150^\circ$ and
$180^\circ$.
With these definitions, the sample contains 686 edge-on systems and 296 face-on systems. The remaining
1{,}745 systems have intermediate
inclinations (see top panel of Figure~\ref{fig:inclination_distribution}).

Figure~\ref{fig:sample_parameters} shows the distributions of many other
parameters in the full sample as well as the face-on and edge-on subsets. In order for a difference in line-broadening between the two subsets to be attributed to a difference in stellar inclination, the intrinsic rotation velocity of the two samples must be the same. While the rotation speed should not depend on the direction from which a system is viewed, differences in detectability may arise due to other astrophysical parameters influencing the photometric or spectroscopic data. If the rotation speed depends on any of these parameters, and if their distributions vary between edge-on and face-on systems, a systematic difference in rotation speed between the samples might still occur. Although the parameter distributions for the edge-on and face-on subsets look similar, Kolmogorov–Smirnov (KS) tests revealed three parameters for which the null hypothesis that the values in the edge-on and face-on subsets are drawn from the same distribution is unlikely ($p<0.05$).

The first case is the effective temperature (p\,=\,0.0088). The edge-on systems show a peak in the temperature distribution near 6150\,K. This may be an effect of discrete sampling used for the parameter inference in DR3, resulting in some parameters not being smoothly distributed. These discrete sampling effects, showing artificial over densities or under densities, are evident in the distributions of several parameters in our sample, as shown in Figure~\ref{fig:sample_parameters}. Regardless of the origin of this difference near 6150\,K, it appears minor and is probably
irrelevant to the subsequent analysis, where
we attempted to control for effective temperature.

The second case is the orbital period, where the face-on sample has more weight
at longer periods than the full sample
(p\,=\,0.02). We hypothesize that this
is because, all other things being equal, astrometric characterization is easier for face-on orbits than edge-on orbits.
For example, the uncertainties in orbital eccentricity and inclination are more strongly correlated for edge-on orbits
compared to face-on orbits.
If this is so, then DR3 would contain more orbital solutions for face-on systems than edge-on systems ---  especially when the period approaches the maximum detectable period, inhibiting detection.
Although we cannot be sure of this explanation, we assume that the difference in period distributions does not matter for our analysis, since \texttt{vbroad} and period are not detectably correlated (see Figure~\ref{fig:sample_parameters}). 

The final and most interesting case of a parameter whose distributions in the edge-on and face-on samples appears statistically different is \texttt{vbroad} (p\,=\,$9.3\times10^{-5}$).
The face-on systems tend to have lower values of \texttt{vbroad} than either the complete sample or the edge-on sample.
Another way to express the difference
between the \texttt{vbroad} distributions
of the edge-on and face-on samples is to examine the mean values and the standard errors in the mean.
For the edge-on systems, $\langle \texttt{vbroad} \rangle_{\mathrm{edge}} = 19.03 \pm 0.70$~km\,s$^{-1}$, while
for the face-on systems, $\langle \texttt{vbroad} \rangle_{\mathrm{face}} = 14.20 \pm 0.73$~km\,s$^{-1}$,
a 4.8-$\sigma$ difference.
In the full sample,
$\langle \texttt{vbroad} \rangle_{\mathrm{all}} = 17.49 \pm 0.32$~km\,s$^{-1}$. 

The overall trends are that the face-on systems show narrower lines than
the edge-on systems, and the edge-on
systems show slightly wider lines than the full sample.
These trends are just what would be expected if the directions of the spin and orbital axes were correlated, and it would be unexpected if the directions
were uncorrelated.

\section{Dependence on effective temperature and eccentricity} 
\label{sec:vbroad_diff}
 
\begin{figure}
  \begin{center}
    \includegraphics[width=1\columnwidth]{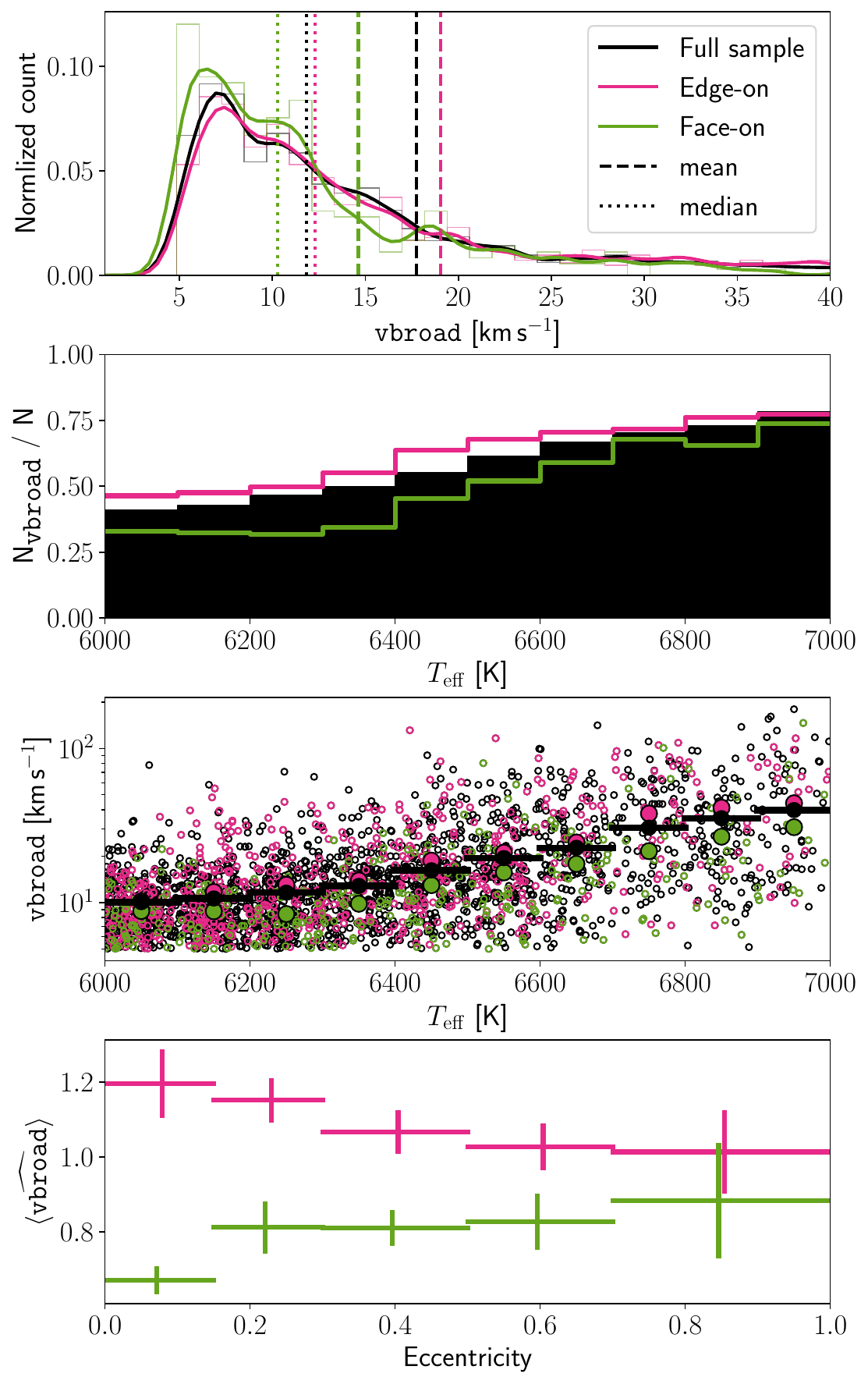}
    \caption{
    \label{fig:basic_result} 
    \textbf{Observed differences in \texttt{vbroad} between face-on and edge-on binaries.}\\ 
    \textit{1st panel:} Smoothed and normalized histograms of the Gaia line-broadening parameter \texttt{vbroad} for all binaries in the sample, as well as for the edge-on or face-on subsamples. 
    \textit{2nd panel:} Fraction of systems for which
    \texttt{vbroad} exceeds 5~km~s$^{-1}$ or, more
    literally, the fraction for which \texttt{vbroad}
    was reported in Gaia DR3. This fraction is shown for the edge-on systems, face-on systems, and the full sample separately, and was split into temperature bins of 100~K. \textit{3rd panel:} \texttt{vbroad} as a function of $T_\mathrm{eff}$. The mean \texttt{vbroad} values of stars in 100~K bins are also shown. Across all temperatures,
    the edge-on sample shows systematically higher values of \texttt{vbroad}, and vice versa for the face-on sample, indicative of spin-orbit alignment.
    \textit{Bottom panel:} The mean value of $\widehat{\texttt{vbroad}}$, the factor by which
    \texttt{vbroad} exceeds the mean value of all
    stars of the same effective temperature.
    This is shown as a function of orbital eccentricity. Horizontal lines indicate the width of the eccentricity bins, and vertical lines indicate the uncertainty. Evidently, nearly-circular binaries exhibit stronger spin-orbit alignment than
    highly eccentric binaries.}
  \end{center}
\end{figure}

The next step in our analysis was to investigate this result in more detail by
testing for any dependence on effective temperature and orbital eccentricity.
Figure~\ref{fig:basic_result} illustrates these investigations.

The top panel shows the
distribution of \texttt{vbroad} for the
full sample and for the two subsamples.
Compared to the edge-on sample, the
face-on sample has more \texttt{vbroad} values below 10~km~s$^{-1}$,
and fewer values of higher line broadening.
The mean and median values of the distributions are also indicated.

The second panel from the top shows a different
line of evidence for systematically
low values of \texttt{vbroad} in the
face-on sample, based on the fact
that the Gaia team chose
not to report \texttt{vbroad} whenever
it was found
to be smaller than 5~km~s$^{-1}$.
Because of this choice,
the fraction of systems for
which \texttt{vbroad} is reported
is an indicator of how many systems have low \texttt{vbroad} values.
The plot shows this fraction as a function
of effective temperature. 
Compared to the full sample, 
the fraction of systems for which \texttt{vbroad} was reported is lower for face-on systems and higher for edge-on systems, across all effective temperatures, indicative of spin-orbit alignment. For this plot, the systems were chosen with the same criteria as in the full sample except for the criteria relating to \texttt{vbroad} and its uncertainty.

The rotation velocities of main-sequence stars have long
been known to be a strong function of effective temperature \citep{Kraft1967}. Hotter and more massive stars rotate
faster, with an especially sharp rise over the temperature range from 6{,}000~K
to 7{,}000~K considered in our study. We might therefore expect to see an even clearer difference in the distributions of \texttt{vbroad}
between face-on and edge-on binaries
if we only compare stars with similar
effective temperatures.
Such a difference can be seen in the third panel from the top in Figure~\ref{fig:basic_result}.
Across the entire
range of effective temperatures, face-on systems have a lower
mean value of \texttt{vbroad} than do 
edge-on systems. To control for this temperature dependence, we defined a star's ``normalized'' value of \texttt{vbroad}, denoted $\widehat{\texttt{vbroad}}$,
to be \texttt{vbroad} divided by the mean \texttt{vbroad} for stars of the same effective temperature. The mean \texttt{vbroad} for a given temperature was calculated by linearly interpolating between the temperature-binned mean \texttt{vbroad} values, represented by black, horizontal lines in the 3rd panel of Figure~\ref{fig:basic_result}. 

The bottom panel of Figure~\ref{fig:basic_result} shows evidence that the difference in the \texttt{vbroad} distributions of face-on and edge-on subsets is most pronounced for binaries
with low eccentricities, and is low or non-existent
for binaries with $e>0.7$. To make this plot, we
grouped the binaries into five bins
according to orbital eccentricity
and computed the mean $\widehat{\texttt{vbroad}}$ of the binaries in each bin, a quantity we denote as $\langle \widehat{\texttt{vbroad}} \rangle$.
We did this entire process
separately for the edge-on systems and
the face-on systems. The plot shows that the primary stars of low-eccentricity,
low-inclination binaries have
systematically narrower spectral lines than
the primary stars in
low-eccentricity, high-inclination
binaries. 
This systematic difference progressively declines 
as binaries with higher eccentricities
are considered, suggesting that eccentricity and spin-orbit alignment are correlated quantities.

\section{Inferring the obliquity distribution} \label{sec:method}

To go beyond testing for statistical
differences and derive quantitative
constraints on the obliquity distribution, we used a forward-modeling approach. We constructed synthetic \texttt{vbroad}
distributions through a Monte Carlo
procedure, starting with a hypothesized obliquity distribution
and simulating the relevant observational effects.
We then determined the ranges of the parameters of the hypothesized obliquity distribution 
that bring the synthetic
distributions into agreement with the measured \texttt{vbroad} distributions. Before this was possible, we needed to
solve two problems:
\begin{itemize}

      \item We needed to calibrate the relationship between \texttt{vbroad} and $v \sin{i}$.
    Section~\ref{sec:calibration} describes our calibration method.
    
    \item We needed a good model for the distribution of stellar rotation velocities as a function of effective temperature. Section~\ref{sec:vrot_model} presents our model.

\end{itemize}

\begin{figure}
  \begin{center}
    \includegraphics[width=1\columnwidth]{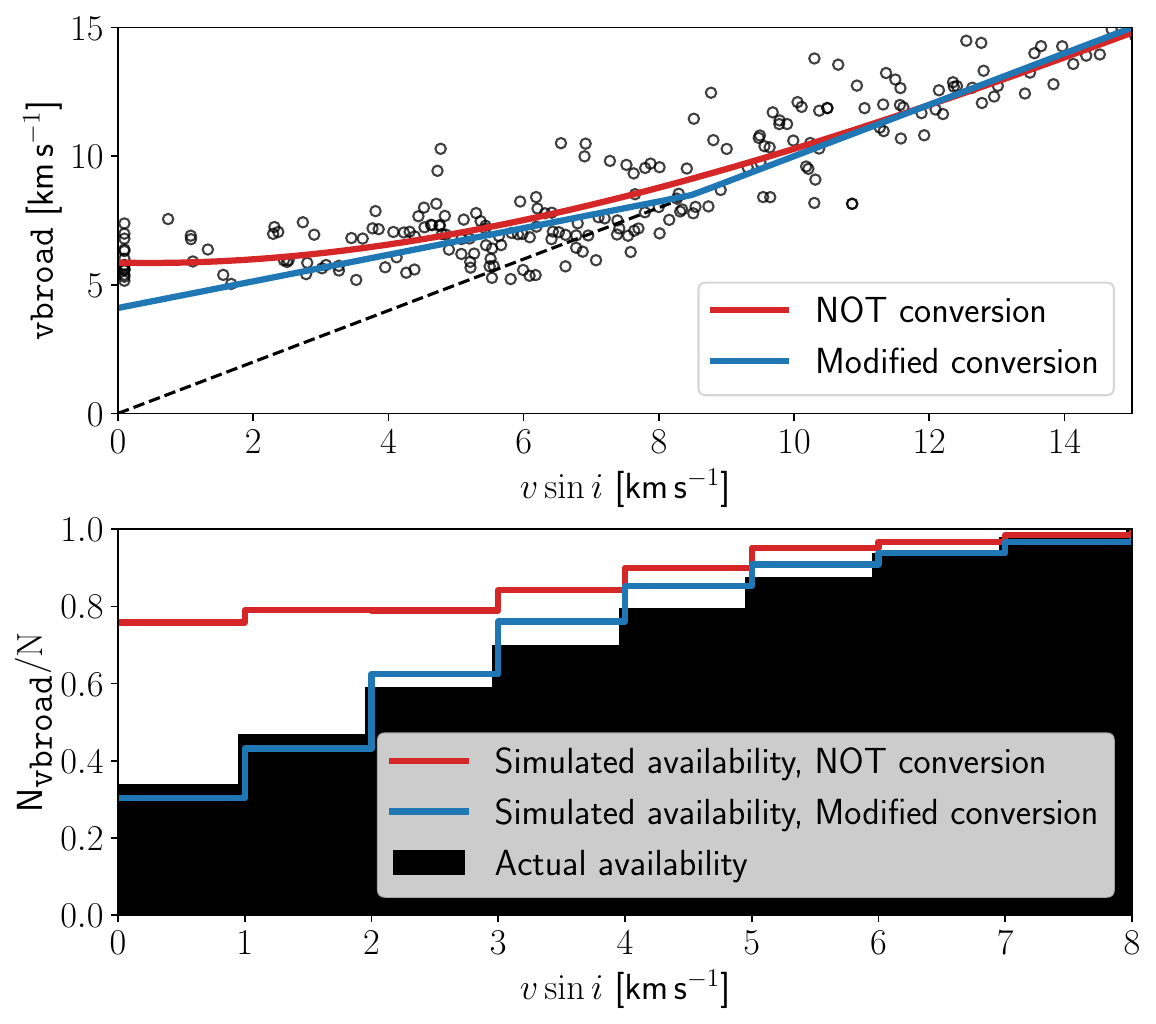}
    \caption{
    \label{fig:FIES_sample} 
    \textbf{Relationship between \texttt{vbroad} and $v \sin i$ based on our FIES calibration sample}. The top panel shows \texttt{vbroad} and $v \sin i$ for 203 stars. The dashed black line is the identity line. The red curve shows a simple fit to the data, and the blue curve is our adopted relationship (see Equation \ref{eq:vbroad_given_vsini}). The black histogram in the bottom panel shows the fraction of all the stars we observed with the NOT for which
    \texttt{vbroad} was reported in Gaia DR3 (and therefore exceeds 5\,km\,s$^{-1}$). The red and blue histograms are based on simulations employing the red and blue calibration curves depicted in the top panel.}
  \end{center}
\end{figure}

\subsection{Relationship between \texttt{vbroad} and $v\sin i$}\label{sec:calibration}

The \texttt{vbroad} parameter in Gaia DR3 was obtained by modeling the observed
spectral lines under the assumptions
that the star is single
and the only
broadening mechanism is rotation \citep{GaiaVbroad2023}.
Other broadening effects such as
turbulent convection and pressure broadening were not modeled. Even in the absence of those other effects, imperfect modeling of instrumental
broadening can cause \texttt{vbroad}
to differ from $v\sin i$.
Additionally, as mentioned above,
whenever \texttt{vbroad} was found to be smaller than 5~km\,s$^{-1}$, it was not
reported in DR3.
We needed a model that predicts, as a function
of $v\sin{i}$,
both the value of \texttt{vbroad} and
the probability that \texttt{vbroad} would be reported in Gaia DR3.

As a starting point,
we conducted spectroscopic observations
of bright F-type stars in order to
measure $v\sin i$ with higher
spectral resolution than the Gaia RVS instrument.
We obtained spectra with a signal-to-noise ratio of about 80 and a spectral
resolution of about 67\,000 (nearly six times higher than the RVS) using the FIES spectrograph at the Nordic Optical Telescope \citep{NOT2010}. We measured $v\sin{i}$ with the iSpec tool \citep{BlancoiSpec2014}. We observed 279 systems for which $v\sin i$ turned out to be below $15\,\mathrm{km}\,{\mathrm{s}^{-1}}$, with results shown in Figure~\ref{fig:FIES_sample}. Of these 279 systems, there were 203 for which \texttt{vbroad} was reported in Gaia DR3. As expected, the frequency with which \texttt{vbroad} was reported is lower for the lower-$v\sin i$ systems; see the black histogram in Figure~\ref{fig:FIES_sample}. The manner in which the frequency declines depends on the relationship between \texttt{vbroad} and $v\sin i$. We took advantage of this fact by adjusting our calibration relationship until the simulations agreed with the observations.
The bottom panel of Figure~\ref{fig:FIES_sample}
shows a comparison between the data and our simulations of the ``availability frequency'' of \texttt{vbroad} as a function of $v\sin i$. Based on these considerations, the formula we adopted to calculate \texttt{vbroad}
(in km~s$^{-1}$) based on $v\sin i$ was:
\begin{equation}
\texttt{vbroad} = \begin{cases} 
4.1\,\left(1 - \frac{x}{8.5}\right), &\leq x \leq 8.5 \\
x, & x > 8.5,
\end{cases}
\label{eq:vbroad_given_vsini}
\end{equation}
where $x$ is $v\sin i$ expressed in km~s$^{-1}$.
This formula is shown as a blue curve in Figure~\ref{eq:vbroad_given_vsini}. The red curve is a polynomial fit between \texttt{vbroad} and $v\sin i$,
which appears to be a good fit but does not
correctly reproduce the fraction of systems
for which $\texttt{vbroad}$ is reported (and therefore
for which it is lower than $5\,\mathrm{km}\,{\mathrm{s}^{-1}}$). We have reported on this calibration procedure for completeness. However, since we could not be sure that
the F-type stars we observed with FIES spanned the
same range of masses and ages as the
primary stars in our sample of astrometric binaries,
and out of concern that stars
in binaries might have systematically
different rotation velocities
than field stars,
we ultimately decided to use the parameters of this FIES-based calibration only as initial guesses for the parameters in the more flexible model described below.

\begin{table}[]
\begin{center}
\begin{tabular}{ll}
Parameter & Value             \\ \hline
$k$       & $0.197\pm 0.003 $  \\
$a_m$     & $0.017\pm 0.0009 $~km\,s$^{-1}$ \\
$a_s$     & $26.0 \pm1.6$~km\,s$^{-1}$ \\
$b_s$     & $21.9\pm0.46$~km\,s$^{-1}$ \\
$c_s$     & $21.4\pm0.78$~km\,s$^{-1}$ 
\end{tabular}
\caption{Parameters describing the overall $v$ distribution in our binary star sample used in Equation~\ref{eq:v_relation}.}
    \label{tab:vTeff_params}
    \end{center}
\end{table}

\subsection{Model for rotation velocity}
 \label{sec:vrot_model}

In our simulations, we needed to assign a rotation velocity to each star. Many ``gyrochronological'' relationships have been established between mass, age, and
rotation velocity, but
none were derived for binaries
with our chosen characteristics.
Instead, we posited a simple stochastic function to describe the
dependence on effective temperature
of both the mean rotation
velocity and the spread in rotation velocities:
\begin{align} \label{eq:v_relation}
    v(T_{\rm eff}) &= v_\mathrm{min}(T_{\rm eff})+ v_\mathrm{scat} (T_\mathrm{eff}) [\mathcal{U}(0,1)^{-k } - 1],
\end{align}
where
\begin{align}
    v_\mathrm{min}(T_{\rm eff}) &= a_m (T_{\rm eff} - 6000\,{\rm K}) \\
    v_\mathrm{scat}(T_{\rm eff}) &= a_s + b_s (T_{\rm eff} -6250\,\mathrm{K}) + c_s (T_{\rm eff} -6250\,\mathrm{K})^2
\end{align}
Here, $\mathcal{U}(0,1)$ is a random
number drawn from a uniform distribution
between 0 and 1, $v_{\mathrm min}$ is the minimum rotation velocity, and $v_{\mathrm scat}$ is the scatter in the velocity distribution. The scatter might arise from the stochastic nature of star formation and evolution as well as variation in other stellar parameters such as surface gravity and metallicity. If the stars strictly obeyed the Skumanich Law ($v \propto \tau^{-1/2}$) and their ages were drawn randomly from a uniform distribution, then $k$ would have a value of 0.5. 

To find the best-fit values for the parameters $k$, $a_m$, $a_s$, $b_s$, and $c_s$, we needed a sample of systems for which the distribution of stellar inclinations is known. For this purpose, we assumed
that the stellar spin axes in our entire sample
of binaries have directions drawn from an isotropic
distribution. This seemed safe because the orbital
inclinations are observed to have a nearly
isotropic distribution, and if so, then the stellar inclinations should also be isotropically distributed, irrespective of the underlying degree of spin-orbit alignment. We also assumed that $v$ depends only on effective temperature, and that a single function $v(T_{\rm eff})$ is applicable to all stars in the sample.

\begin{figure*}
  \begin{center}
    \includegraphics[width=1\textwidth]{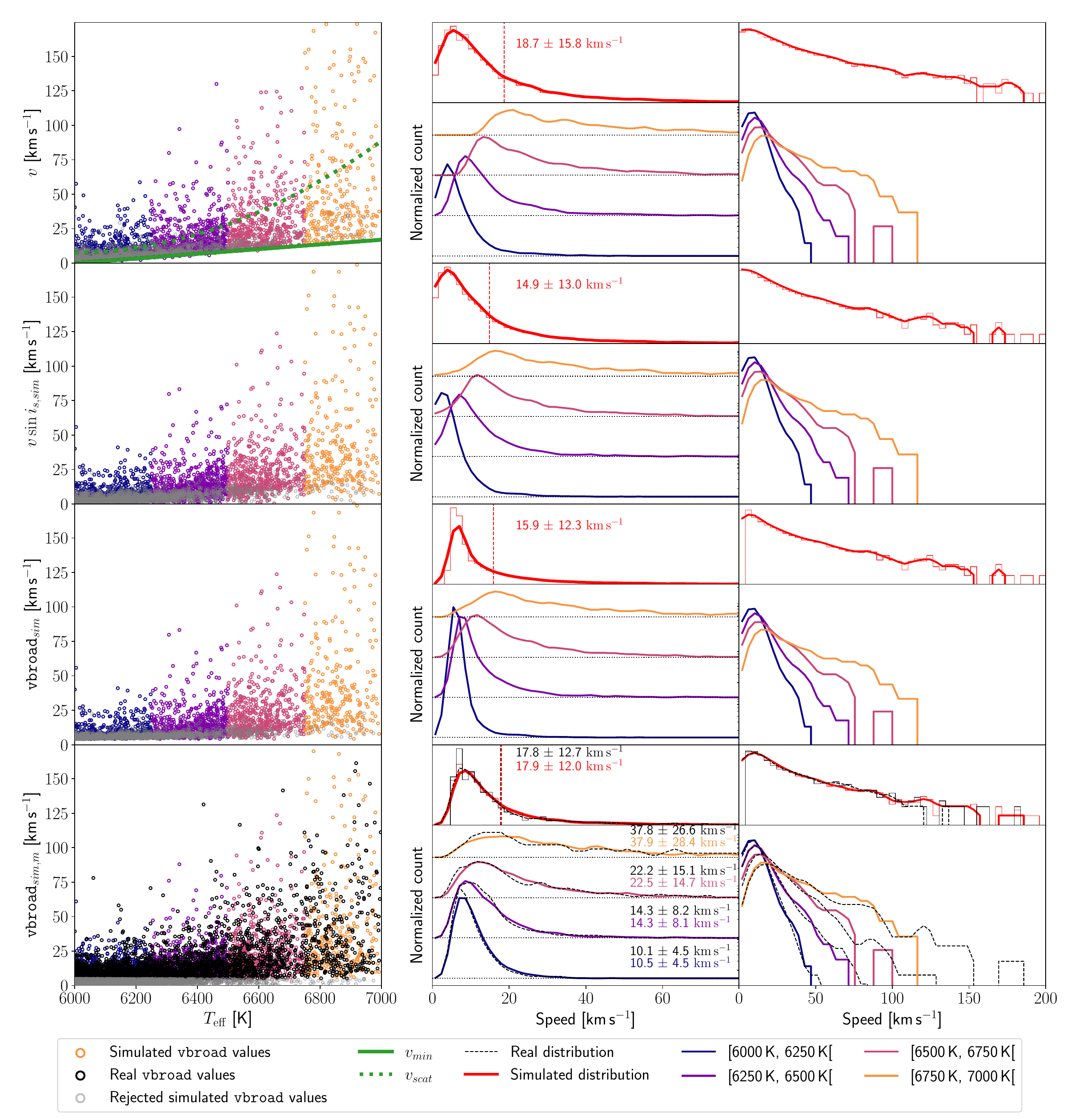}
    \caption{
    \label{fig:simulation} 
    \textbf{Empirical calibration of rotation speed and \texttt{vbroad}.} This $4\times 3$ grid of panels
    illustrates the steps in our model relating $v$, \texttt{vbroad}, and effective temperature.
    The first row shows simulated rotation speeds $v$,
    with the first column displaying the distribution of values, and the second \& third columns showing
    histograms on linear \& logarithmic scales.
    Also shown are the best-fit functions for $v_\mathrm{min}$ and $v_\mathrm{scat}$.
    The second row shows simulated
    $v\sin i$ values, assuming
    isotropic orientations.
    The third row shows the corresponding values of 
    \texttt{vbroad} using our
    conversion formulas.
    The fourth row shows simulated
    measurements of \texttt{vbroad}, taking the statistical uncertainties and the floor of $5$~km\,s$^{-1}$ into account.
    The black points in the lower left
    panel are the real values of
    \texttt{vbroad} in the sample.
    Different colors are used for
    different temperature ranges, as indicated in the legend. 
    }
  \end{center}
\end{figure*}

To assess the agreement between the observed 
and simulated \texttt{vbroad} distributions, we defined a similarity metric based on the means and widths of the distribution of \texttt{vbroad} for the primary stars in each of four $T_\mathrm{eff}$ bins. These bins spanned 250\,K each. Specifically, the following statistic was minimized:
\begin{align}
    \chi^2 = &\sum_{i=1}^4 \left( \frac{\mu_\mathrm{sim,T_{i}} - \mu_\mathrm{real,T_i}}{\mathrm{SE_\mu}}  \right)^2 +\\ &\sum_{i=1}^4 \left( \frac{\sigma_\mathrm{sim,T_i} - \sigma_\mathrm{real,T_i}}{\mathrm{SE_\sigma}}  \right)^2,
\end{align}
where $\mathrm{T_i}$ refers to the $i$th temperature bin,\footnote{For clarity we dropped the subscript "eff" for temperature in these equations.}
$\mu$ is the mean value of \texttt{vbroad},
and $\mathrm{SE}_\mu$ and $\mathrm{SE}_\sigma$ refer to the standard error of the mean and standard deviations, respectively. 
The binned \texttt{vbroad} distributions are shown in different colors in Figure~\ref{fig:simulation}. The optimal parameters were found by minimizing $\chi^2$, and the parameter uncertainties were found by
perturbing each parameter away from the optimal value
until $\chi^2$ increased by one unit.

The optimal parameter values are given in Table~\ref{tab:vTeff_params}. The resulting functions for $v_{\mathrm scat}$ and $v_{\mathrm min}$ are also shown in green in the top left panel of Figure~\ref{fig:simulation}.

\begin{figure}
  \begin{center}
    \includegraphics[width=1\columnwidth]{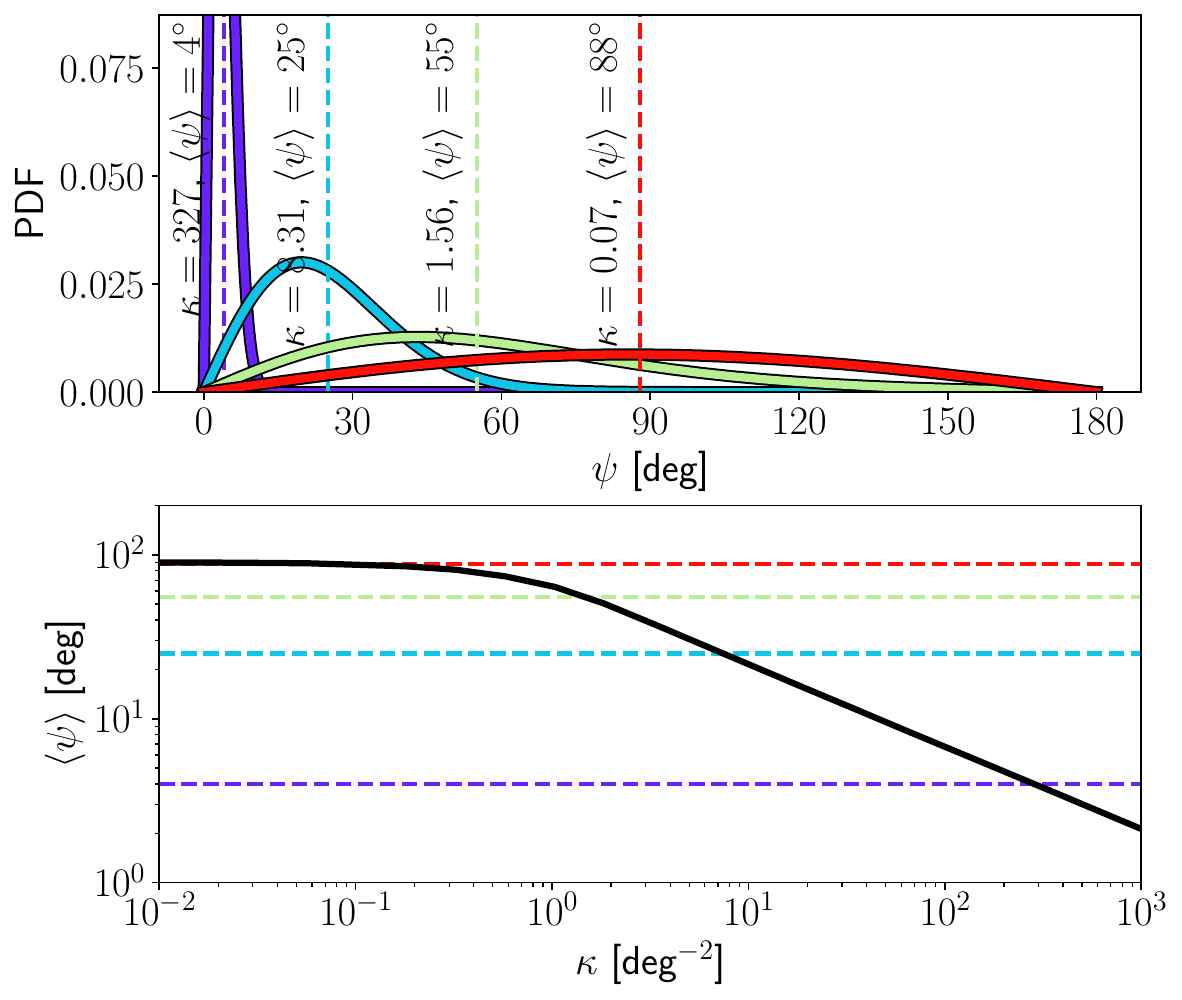}
    \caption{
    \label{fig:kappa_meanpsi} 
    \textbf{von-Mises Fisher distribution.}\\
    {\it} Top: The VMF distribution is exemplified for four different concentrations, quantified by both the concentration $\kappa$ and the mean of $\psi$, $\langle \psi \rangle$. The dashed lines represent the $\langle \psi \rangle$ values. {\it Bottom:} How $\kappa$ and $\langle \psi \rangle$ relate to one another superimposed on the four values of $\langle \psi \rangle$ from the top panel. 
    }
  \end{center}
\end{figure}

\subsection{Obliquity distribution and resulting stellar inclinations} 
\label{sec:Chosen obliquity distribution}

As a simple parametric model for the distribution of stellar obliquities, we decided to follow \cite{fabrycky_winn2009} and use the von-Mises Fisher (VMF) distribution,
\begin{equation}
    p (\psi) = \frac{\kappa}{2 \sinh{\kappa}} \exp(\kappa \cos{\psi})\sin \psi.
\end{equation}
This distribution resembles a 2-d Gaussian distribution wrapped around a sphere. The degree of spin-orbit alignment is quantified by the concentration parameter $\kappa$. However, since the relation between $\kappa$ and the width of the distribution is not
straightforward, we chose to parameterize our results in terms of the mean of the spin-orbit angle $\psi$ of the VMF distribution. For concentrated distributions, with $\kappa \gg 1$, the mean obliquity $\langle \psi \rangle \approx \kappa^{-1/2}\sqrt{\pi/2}$. For broad distributions, for which $\kappa$ approaches $0$, $\langle \psi \rangle \rightarrow 90^\circ$. The general relationship between $\kappa$ and $\langle \psi \rangle$ was determined numerically with a Monte Carlo approach. A large number of $\psi$ values were drawn from a VMF distribution with concentration parameter $\kappa$, the mean was calculated and recorded,
and the procedure was repeated for many choices of
$\kappa$. The results are shown in the bottom panel of Figure~\ref{fig:kappa_meanpsi}, relating $\kappa$ to $\langle \psi \rangle$. In the top panel of the same figure, the VMF distribution is illustrated for four values of $\langle \psi \rangle$.

To specify the direction of the spin axis
in three dimensions, $\psi$ must be supplemented
with an azimuthal angle $\Omega$ that specifies the
direction of the component of the spin axis
that is perpendicular to the orbit, following \cite{fabrycky_winn2009}: We assumed that $\Omega$ is uniformly distributed
between $0^\circ$ and $360^\circ$. Once
$\psi$ and $\Omega$ are chosen,
we can calculate the sky-projected spin-orbit angle $\lambda$ and the stellar inclination $i$ using Equations~11 and 8 of \cite{fabrycky_winn2009}:
\begin{align*}
    \lambda  & =  \mathrm{arctan} \left( \frac{\sin{\psi}  \sin{\Omega_{sim}}}{\cos{\psi}\sin{i_o} + \sin{\psi}\cos{\Omega}\cos{i_o}} \right) \\
    \sin{i} &= \frac{\sin{\psi} \sin{\Omega}}{\sin\lambda}.
\end{align*}

\subsection{Summary of simulation
methodology}\label{sec:sim}

We now summarize the method by which
simulated distributions of \texttt{vbroad}
were created and compared with observations. Ultimately, the goal was to compare the observed distributions of $\langle \widehat{\texttt{vbroad}}\rangle$ of the edge-on and face-on binaries with those
of synthetic datasets that were created assuming
different degrees of spin-orbit
alignment. First, we determined the parameters of an empirical relationship between $v(T_\mathrm{eff})$ and \texttt{vbroad}$(v\sin{i})$ by assuming that the primary stars in the entire sample are isotropically oriented. This was described in Sections~\ref{sec:vrot_model} and \ref{sec:calibration}. Then, we postulated a particular obliquity distribution (parameterized by $\langle \psi \rangle$) and 
assigned a simulated value of $\texttt{vbroad}$ to each star in a sample with the following steps:
\begin{enumerate}
    \item Assign $i_o$ and $T_{\mathrm{eff}}$ values and simulate measurements of these quantities based on the reported observational uncertainties. 
    \item Draw $\psi$ from a VMF distribution with the assumed value of $\langle \psi \rangle$.
    \item Calculate the stellar inclination $i$ based on the assigned $i_o$ and $\psi$ values.
    \item Assign a rotation velocity $v$ using the relationship $v(T_\mathrm{eff})$ specified in Equation~\ref{eq:v_relation}.
    \item Calculate $v\sin i$ based on $v$ and $i$.
    \item Convert $v\sin{i}$ into \texttt{vbroad} using the calibrated
    relationship specified in Equation~\ref{eq:vbroad_given_vsini}.
    \item Simulate a measurement of \texttt{vbroad} by 
    drawing a value from a normal distribution
    centered at the calculated value
    of \texttt{vbroad}, and with a width
    equal to the uncertainty reported in Gaia DR3. The distribution was truncated
    at 0\,km\,s$^{-1}$ to prevent negative values from being drawn.
    \item Whenever the result for
    \texttt{vbroad} was below 5\,km\,s$^{-1}$,
    we repeated all the steps from the beginning to try again, since Gaia DR3 only reported \texttt{vbroad} when it was
    found to exceed 5\,km\,s$^{-1}$.
\end{enumerate}
The four rows of Figure~\ref{fig:simulation} show snapshots in the production process of a simulated sample at steps 4, 5, 6, and 8. 

\subsection{Inference of mean obliquity}
\label{sec:inferring_alignment}

\begin{figure*}
  \begin{center}
    \includegraphics[width=1\textwidth]{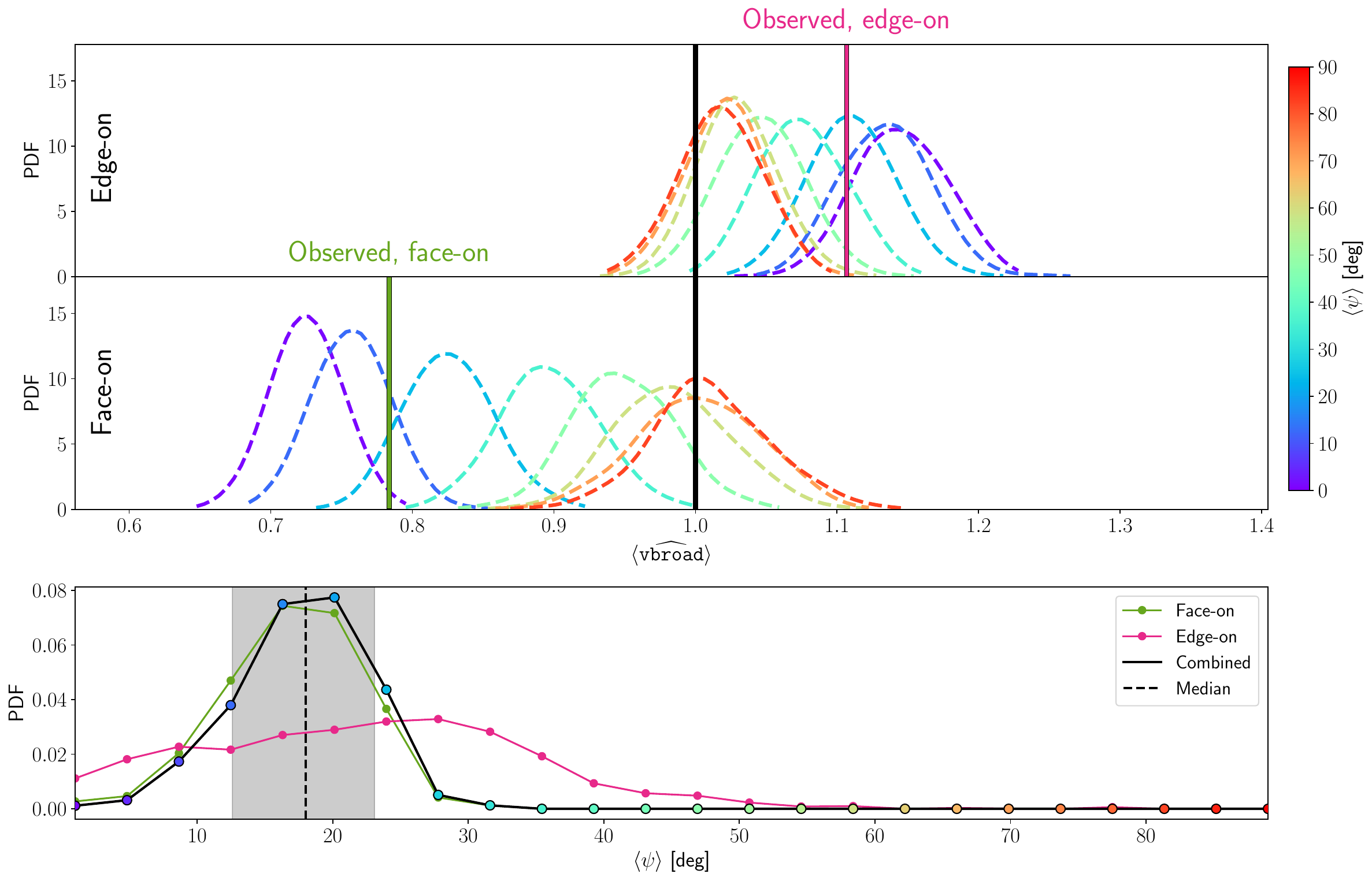}
    \caption{
    \label{fig:posteriors} 
    \textbf{Edge- and face-on alignment.}
    {\it Top two panels:} Each colored distribution shows the probability of making a certain measurement of $\langle\widehat{\texttt{vbroad}}\rangle$, given an assumed degree of spin-orbit alignment (quantified by $\langle \psi \rangle$). Vertical lines mark
    the observed values of
    $\langle\widehat{\texttt{vbroad}}\rangle$
    across all eccentricities, for the edge-on
    sample (top) and the face-on sample (middle).
 {\it Bottom panel:} The posterior PDF of $\langle \psi \rangle$
    derived from the edge-on (pink) and face-on (green) samples, as well as their combination (black). The results of analyzing the two subsets are compatible, and the face-on sample gives
    tighter constraints. 
    }
  \end{center}
\end{figure*}

To infer the spin-orbit alignment of the edge-on and face-on samples, we created synthetic samples for 25 different choices of $\langle \psi \rangle$ spaced evenly between
$0^\circ$ and $90^\circ$. The randomness in the simulations is due to several factors: the measurement uncertainties in $i_{\mathrm o}$, \texttt{vbroad}, and $T_\mathrm{eff}$; the stochastic relationship between $v$ and $T_{\rm eff}$; and the random draws of obliquities
from the posited VMF distribution. Three hundred synthetic samples were created for each of the 25 values of $\langle \psi \rangle$. 
The simulated value of $\langle \widehat{\texttt{vbroad}} \rangle$ was calculated for each sample, and their
distribution was taken to be the probability
density for $\langle \widehat{\texttt{vbroad}} \rangle$ for the given value of $\langle \psi \rangle$. See the dashed and colored lines in  Figure~\ref{fig:posteriors}. We used these probability density functions to calculate the posterior probability density function of $\langle \psi \rangle$ given the observed value of $\langle \widehat{\texttt{vbroad}} \rangle$.

\section{Results} \label{sec:results}

First, we analyzed the full sample, regardless of orbital eccentricity. We expect the spin-orbit distributions of the edge-on and face-on binaries to be indistinguishable, because there should be no dependence of a system's intrinsic geometry on the direction from which it is viewed.
Figure~\ref{fig:posteriors} displays the results. In the top two panels, the observed values of $\langle \texttt{vbroad} \rangle$ are shown as vertical lines. Alongside them are dashed colored curves that were constructed from the synthetic versions of the same samples. Each color represents the probability of measuring $\langle \texttt{vbroad} \rangle$ for a given value of $\langle \psi \rangle$. The bottom panel shows the posterior PDFs for $\langle \psi \rangle$ as informed by the face-on binaries (green), edge-on binaries (pink), and the entire sample (black). The results from the edge-on and face-on samples are consistent, as expected, and indicate $\langle \psi \rangle =18 \pm 5^\circ$. Thus, the binaries as a whole show a moderate degree of spin-orbit alignment.

Next, motivated by the experiments on eccentricity described in Section~\ref{sec:vbroad_diff} and displayed in Figure~\ref{fig:basic_result}, we divided the sample into 5 subsamples according to eccentricity, and performed the preceding analysis on each subsample. The eccentricity ranges that defined the bins were $[0-0.15),[ 0.15-0.30),[0.30-0.50),[0.50-0.70)$, and $[0.70-1]$. The number of binaries in each eccentricity bin was 568, 729, 780, 474, and 176, respectively.

\begin{figure*}
  \begin{center}
    \includegraphics[width=1\textwidth]{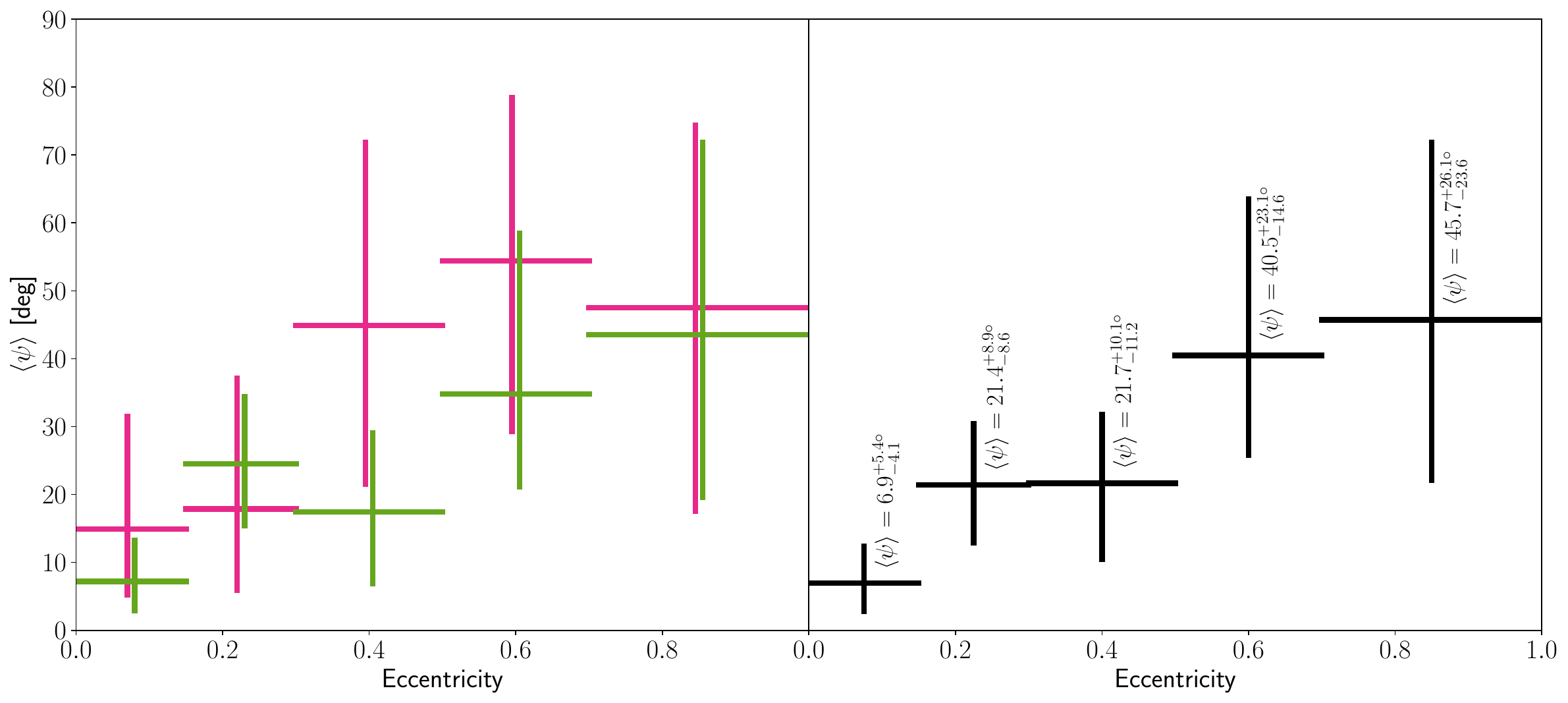}
    \caption{
    \label{fig:psi_e} 
    \textbf{High obliquities are associated with high orbital eccentricity.}
    The results for the spin-orbit alignment of our binary star sample are shown as a function of eccentricity in five bins. The eccentricity ranges are indicated by horizontal bars and vertical bars represent the one-sigma confidence intervals for $\langle \psi \rangle$. Pink, green and black signify edge-on, face-on, and combined results, respectively. The green error bars were displaced slightly to the right to avoid overlapping with the pink error bars.
    }
  \end{center}
\end{figure*}

Figure~\ref{fig:psi_e} shows the results, combining the information from the face-on and edge-on systems for the five eccentricity bins. Systems with eccentricities below 0.15 have an obliquity distribution consistent with perfect alignment. As eccentricity increases, the obliquity dispersion also increases. Systems with eccentricities above 0.7 have a mean obliquity of $\langle \psi \rangle = 46_{-24}^{+26^\circ}$.

\subsection{Caveats and robustness of the result} \label{subsec:caveats}

A potential source of systematic error is the unmodelled effect of light from the secondary star in each binary on the determination of \texttt{vbroad} from the RVS spectra. Any bias in \texttt{vbroad} due to the secondary star might depend on orbital inclination, because the line-of-sight component of the orbital motion is higher for an edge-on system than for a face-on system. In addition, a significant contribution of light from the secondary star would probably lead to an increase in the reported uncertainty of \texttt{vbroad}, because
the uncertainty reported in Gaia DR3 is based on the standard deviation of the time series of individual \texttt{vbroad} measurements. We did not find any systematic difference in the distribution of \texttt{vbroad} uncertainties between the face-on and edge-on systems. Regardless, in an attempt to exclude the most problematic systems, we required a stringent upper limit of 10~km\,s$^{-1}$ on the uncertainty of \texttt{vbroad}.

We also need to remind ourselves that the stellar inclination derived from $v\sin i$ is subject to a discrete degeneracy; we cannot
distinguish $i$ from $180^\circ-i$. If the true obliquity distribution of the binary systems is well approximated by the VMF distribution, then our results are probably not affected much by this degeneracy. However, if the real probability distribution does not decline monotonically with $\psi$
and has an excess of retrograde systems, then
the degeneracy might be serious.
For example, if the true obliquity
distribution were bimodal with peaks near both $\psi = 0^\circ$ and $\psi = 180^\circ$, then our method would falsely infer that the distribution is only concentrated
near $0^\circ$. The degeneracy in the stellar inclination can be broken, in principle, through interferometric observations (see \citealt{AlbrechtDawsonWinn2022} and references therein). 

Finally, we note that the results of analyzing the edge-on and face-on subsets are consistent with each other. However, the most powerful constraints come
from the face-on subset with low-eccentricity orbits.
This is because the main observational implication of spin-orbit alignment is that $\sin i$ should be systematically low for face-on systems. For face-on systems, $i$ is low and $\sin i \approx i$. Therefore any change in $i$ leads a proportional change in $\sin i$. For edge-on systems, $i\approx \pi/2$ and
$\sin i \approx 1 - \alpha^2/2$ where
$\alpha = \pi/2 - i$.
Therefore, a change in $i$ leads
to only a second-order change in $\sin i$.
The effects of this flattening of the sine function with $i$ is familiar
from similar studies
of stars with transiting planets
\citep[see, e.g.,][]{Winn+2017}, and
causes face-on samples to bear
most of the statistical power
when constraining the well-aligned systems.
The advantage of face-on systems over edge-on systems
is reduced for samples with larger spin-orbit misalignments, and reversed for nearly polar orbits.

\section{Discussion} \label{sec:discussion}

We have found that in binaries
with F-type primaries and periods between
50 and 1000 days, spin-orbit misalignment of the primary star is associated
with high orbital eccentricity.
Dissipative tidal interactions between the two stars would naturally lead to
both spin-orbit alignment and orbital circularization. However, with periods exceeding 50 days, we would not expect
ongoing tidal interactions to be significant \cite[see, e.g.,][]{JustesenAlbrecht2021,Bashi+2023}. Tidal dissipation would have been more rapid early in the stars' lives when they were larger and still contracting onto the main sequence. But the absence of a detectable dependence of \texttt{vbroad} (and by extension, $v$) on orbital period suggests that alignment and synchronization of the stellar spin is not the dominant factor in shaping the stellar obliquities in our sample.

Instead, the eccentricity-obliquity correlation may have been imprinted during formation of these binaries (see \citealp{Offner+2023} for an extensive discussion of stellar binary formation).
To begin, while in-situ formation of stellar binaries via the fragmentation of turbulent molecular cores has been invoked to explain eccentricity statistics of wider stellar binaries \citep{2023xu_superthermal}, such a process is not thought to efficiently form stellar binaries with separations $\lesssim 10^2$ au \citep[e.g.][]{2017Guszejnov_turb100au, Offner+2023}.
In order to form binaries with periods comparable to those in our sample, three classes of theories can be invoked: capture from an unbound state, gas-driven migration from an initially wider orbit, and secular processes induced by a distant tertiary companion \citep{Offner+2023}.
In the first case, sufficiently dense stellar environments can yield compact binaries (separations $\lesssim$ au) via gravitational capture and even partner exchange \citep[e.g.][]{2017dorval}, though such processes may not be as efficient in gas-rich environments \citep[e.g.][]{2019wall_capture}.
Naively, such dynamically violent processes should not produce many circular binaries with low obliquities, though they may be responsible for the eccentric, misaligned binaries in our sample.
In the second case, gas-driven migration has classically been thought to result in binary coalescence \citep{1996arty_circumbinarydisk}, though recent studies have concluded that the evolution of the binary separation and eccentricity may be more complex than previously thought (see e.g.\ \citealp{2023lai} for a review), and a direct theoretical prediction for the binary properties at the end of such a process is uncertain.
Finally, in the third scenario, a tertiary companion may induce oscillations of the eccentricity in the binary system and the mutual inclination between the binary system and the tertiary system via the von Zeipel-Lidov-Kozai effect \citep{vonzeipel, lidov1962, kozai1962}. Such oscillations, in conjunction with standard theories of tidal dissipation, can efficiently form binaries with periods $\lesssim 10$ days \citep{fabrycky2007}, but are unlikely to form the wider binaries in our sample, where tidal dissipation is inefficient.

In summary, while the formation of binaries with the orbital periods considered in our sample is still not well understood, the most promising mechanisms involve dynamical processes that should often lead to eccentric and misaligned binaries.

Separately, a possible suspect for the observed correlation is the interaction of a binary with a third star or multiple stars in the birth cluster after its formation.
Starting from a low-obliquity, low-eccentricity state, such gravitational interactions would tend to excite both the binary's orbital eccentricity \citep[e.g.][]{1996HeggieRasio} and its orbital inclination \citep[e.g.][]{2021rodetsulai}. Close encounters with other stars or multiple star systems could lead to scattering events that change the velocity vector of one star in the binary and therefore change both orbital eccentricity as well as the orbital plane. 

Under this hypothesis, one might expect that more closely bound systems with shorter orbital periods would be more protected and would display lower obliquities and eccentricities. We do not see a clear trend with period for eccentricity and obliquities. 
Additionally, for the short periods of binaries considered here, the rate of encounters sufficiently strong to affect the orbit \citep[$\sim 1\;\mathrm{Gyr^{-1}}$, e.g.][]{2021rodetsulai} is likely too slow to affect the binary orbit within characteristic birth cluster lifetimes \citep[$\sim 10\;\mathrm{Myr}$, e.g.][]{2005lamers_lifetimes}, though this comparison depends on the highly uncertain properties of stellar birth clusters and the age of the cluster at the time of any dynamical interaction.

Our results can also be compared to measurements of the obliquities of single stars with known planetary systems.  It appears as if the alignment of the low-eccentricity binaries in our sample is comparable to that of the most well-aligned hot Jupiter systems \citep{AlbrechtDawsonWinn2022} as well as most of the stars hosting compact multi-transiting planetary systems (Kundstrup et al., under review). It also seems to be consistent with the good alignment of the Sun with respect to the orbital planes of the planets in the Solar System.

Future studies might be able to test if the above processes or other processes lead naturally to the observed distribution of spin-orbit angles and its dependence on eccentricity. This will be possible not only because the number of systems is expected to increase with DR4. The longer time baseline (DR3 included 34 months of data) will allow for longer period systems to be included. DR4 will also contain epoch astrometry and RVS spectra. In particular, the epoch RVS data allows for the inclusion of SB2 binaries, thereby extending the mass ratio range. The DR4 data will also be sensitive to lower mass secondaries. Future studies might also test for any age dependency. Using samples like the one presented by \cite{Hwang2023} will also allow the study of systems with significantly larger separations and allow an investigation of the role multiplicity.
Higher spectral resolution for lower temperature systems should allow us to probe systems with stars outside the spectral range probed here. For example APOGEE \citep{SDSS17} and GALAH \citep{GALAH+} spectra have more than twice the resolution of Gaia RVS spectra, whereas single object high resolution echelle spectrographs have resolutions of about an order of magnitude higher. Finally, our methodology could be further developed including e.g.\ ages and metallicity into the rotation modeling as well as employing a hierarchical approach for obliquity and model parameter inference.

\begin{acknowledgments}
This research benefited from using the Gaiachatbot at \url{https://gaiachat.streamlit.app/} provided by Veridicity. 

This work has made use of data from the European Space Agency (ESA) mission {\it Gaia} (\url{https://www.cosmos.esa.int/gaia}), processed by the {\it Gaia} Data Processing and Analysis Consortium (DPAC, \url{https://www.cosmos.esa.int/web/gaia/dpac/consortium}). Funding for the DPAC has been provided by national institutions, in particular the institutions participating in the {\it Gaia} Multilateral Agreement.

We acknowledge the support from the Danish Council for Independent Research through a grant, No.2032-00230B.

This work was supported by a research grant (42101) from VILLUM FONDEN as well as The Independent Research Fund Denmark’s Inge Lehmann program (grant agreement No. 1131-00014B).

This project has received funding from the European Union’s Horizon 2020 research and innovation programme under grant agreement No 101004719 (ORP: OPTICON RadioNet Pilot)

S.H.A.\ thanks the Department of Astrophysical Sciences at Princeton University and the Action Station chef of the faculty lunch room at Prospect House for their hospitality during a sabbatical stay during which part of the work for this project was conducted.
\end{acknowledgments}

\vspace{5mm}
\facilities{NOT \citep{NOT2010}, Gaia \citep{GaiaMission}}

\software{iSpec \citep{BlancoiSpec2014}}

\bibliography{references}{}

\begin{thebibliography}{}
\expandafter\ifx\csname natexlab\endcsname\relax\def\natexlab#1{#1}\fi
\providecommand{\url}[1]{\href{#1}{#1}}
\providecommand{\dodoi}[1]{doi:~\href{http://doi.org/#1}{\nolinkurl{#1}}}
\providecommand{\doeprint}[1]{\href{http://ascl.net/#1}{\nolinkurl{http://ascl.net/#1}}}
\providecommand{\doarXiv}[1]{\href{https://arxiv.org/abs/#1}{\nolinkurl{https://arxiv.org/abs/#1}}}

\bibitem[{{Abdurro'uf} {et~al.}(2022){Abdurro'uf}, {Accetta}, {Aerts}, {Silva Aguirre}, {Ahumada}, {Ajgaonkar}, {Filiz Ak}, {Alam}, {Allende Prieto}, {Almeida}, {Anders}, {Anderson}, {Andrews}, {Anguiano}, {Aquino-Ort{\'\i}z}, {Arag{\'o}n-Salamanca}, {Argudo-Fern{\'a}ndez}, {Ata}, {Aubert}, {Avila-Reese}, {Badenes}, {Barb{\'a}}, {Barger}, {Barrera-Ballesteros}, {Beaton}, {Beers}, {Belfiore}, {Bender}, {Bernardi}, {Bershady}, {Beutler}, {Bidin}, {Bird}, {Bizyaev}, {Blanc}, {Blanton}, {Boardman}, {Bolton}, {Boquien}, {Borissova}, {Bovy}, {Brandt}, {Brown}, {Brownstein}, {Brusa}, {Buchner}, {Bundy}, {Burchett}, {Bureau}, {Burgasser}, {Cabang}, {Campbell}, {Cappellari}, {Carlberg}, {Wanderley}, {Carrera}, {Cash}, {Chen}, {Chen}, {Cherinka}, {Chiappini}, {Choi}, {Chojnowski}, {Chung}, {Clerc}, {Cohen}, {Comerford}, {Comparat}, {da Costa}, {Covey}, {Crane}, {Cruz-Gonzalez}, {Culhane}, {Cunha}, {Dai}, {Damke}, {Darling}, {Davidson}, {Davies}, {Dawson}, {De Lee}, {Diamond-Stanic}, {Cano-D{\'\i}az}, {S{\'a}nchez},
  {Donor}, {Duckworth}, {Dwelly}, {Eisenstein}, {Elsworth}, {Emsellem}, {Eracleous}, {Escoffier}, {Fan}, {Farr}, {Feng}, {Fern{\'a}ndez-Trincado}, {Feuillet}, {Filipp}, {Fillingham}, {Frinchaboy}, {Fromenteau}, {Galbany}, {Garc{\'\i}a}, {Garc{\'\i}a-Hern{\'a}ndez}, {Ge}, {Geisler}, {Gelfand}, {G{\'e}ron}, {Gibson}, {Goddy}, {Godoy-Rivera}, {Grabowski}, {Green}, {Greener}, {Grier}, {Griffith}, {Guo}, {Guy}, {Hadjara}, {Harding}, {Hasselquist}, {Hayes}, {Hearty}, {Hern{\'a}ndez}, {Hill}, {Hogg}, {Holtzman}, {Horta}, {Hsieh}, {Hsu}, {Hsu}, {Huber}, {Huertas-Company}, {Hutchinson}, {Hwang}, {Ibarra-Medel}, {Chitham}, {Ilha}, {Imig}, {Jaekle}, {Jayasinghe}, {Ji}, {Johnson}, {Jones}, {J{\"o}nsson}, {Katkov}, {Khalatyan}, {Kinemuchi}, {Kisku}, {Knapen}, {Kneib}, {Kollmeier}, {Kong}, {Kounkel}, {Kreckel}, {Krishnarao}, {Lacerna}, {Lane}, {Langgin}, {Lavender}, {Law}, {Lazarz}, {Leung}, {Leung}, {Lewis}, {Li}, {Li}, {Lian}, {Liang}, {Lin}, {Lin}, {Lin}, {Lintott}, {Long}, {Longa-Pe{\~n}a}, {L{\'o}pez-Cob{\'a}}, {Lu},
  {Lundgren}, {Luo}, {Mackereth}, {de la Macorra}, {Mahadevan}, {Majewski}, {Manchado}, {Mandeville}, {Maraston}, {Margalef-Bentabol}, {Masseron}, {Masters}, {Mathur}, {McDermid}, {Mckay}, {Merloni}, {Merrifield}, {Meszaros}, {Miglio}, {Di Mille}, {Minniti}, {Minsley}, {Monachesi}, {Moon}, {Mosser}, {Mulchaey}, {Muna}, {Mu{\~n}oz}, {Myers}, {Myers}, {Nadathur}, {Nair}, {Nandra}, {Neumann}, {Newman}, {Nidever}, {Nikakhtar}, {Nitschelm}, {O'Connell}, {Garma-Oehmichen}, {Luan Souza de Oliveira}, {Olney}, {Oravetz}, {Ortigoza-Urdaneta}, {Osorio}, {Otter}, {Pace}, {Padilla}, {Pan}, {Pan}, {Parikh}, {Parker}, {Peirani}, {Pe{\~n}a Ram{\'\i}rez}, {Penny}, {Percival}, {Perez-Fournon}, {Pinsonneault}, {Poidevin}, {Poovelil}, {Price-Whelan}, {B{\'a}rbara de Andrade Queiroz}, {Raddick}, {Ray}, {Rembold}, {Riddle}, {Riffel}, {Riffel}, {Rix}, {Robin}, {Rodr{\'\i}guez-Puebla}, {Roman-Lopes}, {Rom{\'a}n-Z{\'u}{\~n}iga}, {Rose}, {Ross}, {Rossi}, {Rubin}, {Salvato}, {S{\'a}nchez}, {S{\'a}nchez-Gallego}, {Sanderson}, {Santana
  Rojas}, {Sarceno}, {Sarmiento}, {Sayres}, {Sazonova}, {Schaefer}, {Schiavon}, {Schlegel}, {Schneider}, {Schultheis}, {Schwope}, {Serenelli}, {Serna}, {Shao}, {Shapiro}, {Sharma}, {Shen}, {Shetrone}, {Shu}, {Simon}, {Skrutskie}, {Smethurst}, {Smith}, {Sobeck}, {Spoo}, {Sprague}, {Stark}, {Stassun}, {Steinmetz}, {Stello}, {Stone-Martinez}, {Storchi-Bergmann}, {Stringfellow}, {Stutz}, {Su}, {Taghizadeh-Popp}, {Talbot}, {Tayar}, {Telles}, {Teske}, {Thakar}, {Theissen}, {Tkachenko}, {Thomas}, {Tojeiro}, {Hernandez Toledo}, {Troup}, {Trump}, {Trussler}, {Turner}, {Tuttle}, {Unda-Sanzana}, {V{\'a}zquez-Mata}, {Valentini}, {Valenzuela}, {Vargas-Gonz{\'a}lez}, {Vargas-Maga{\~n}a}, {Alfaro}, {Villanova}, {Vincenzo}, {Wake}, {Warfield}, {Washington}, {Weaver}, {Weijmans}, {Weinberg}, {Weiss}, {Westfall}, {Wild}, {Wilde}, {Wilson}, {Wilson}, {Wilson}, {Wolf}, {Wood-Vasey}, {Yan}, {Zamora}, {Zasowski}, {Zhang}, {Zhao}, {Zheng}, {Zheng}, \& {Zhu}}]{SDSS17}
{Abdurro'uf}, {Accetta}, K., {Aerts}, C., {et~al.} 2022, \apjs, 259, 35, \dodoi{10.3847/1538-4365/ac4414}

\bibitem[{{Albrecht} {et~al.}(2007){Albrecht}, {Reffert}, {Snellen}, {Quirrenbach}, \& {Mitchell}}]{albrecht2007}
{Albrecht}, S., {Reffert}, S., {Snellen}, I., {Quirrenbach}, A., \& {Mitchell}, D.~S. 2007, \aap, 474, 565, \dodoi{10.1051/0004-6361:20077953}

\bibitem[{{Albrecht} {et~al.}(2009){Albrecht}, {Reffert}, {Snellen}, \& {Winn}}]{albrecht2009}
{Albrecht}, S., {Reffert}, S., {Snellen}, I.~A.~G., \& {Winn}, J.~N. 2009, \nat, 461, 373, \dodoi{10.1038/nature08408}

\bibitem[{{Albrecht} {et~al.}(2013){Albrecht}, {Setiawan}, {Torres}, {Fabrycky}, \& {Winn}}]{albrecht2013}
{Albrecht}, S., {Setiawan}, J., {Torres}, G., {Fabrycky}, D.~C., \& {Winn}, J.~N. 2013, \apj, 767, 32, \dodoi{10.1088/0004-637X/767/1/32}

\bibitem[{{Albrecht} {et~al.}(2011){Albrecht}, {Winn}, {Carter}, {Snellen}, \& {de Mooij}}]{Albrecht+2011}
{Albrecht}, S., {Winn}, J.~N., {Carter}, J.~A., {Snellen}, I. A.~G., \& {de Mooij}, E. J.~W. 2011, \apj, 726, 68, \dodoi{10.1088/0004-637X/726/2/68}

\bibitem[{{Albrecht} {et~al.}(2014{\natexlab{a}}){Albrecht}, {Winn}, {Torres}, {Fabrycky}, {Setiawan}, {Gillon}, {Jehin}, {Triaud}, {Queloz}, {Snellen}, \& {Eggleton}}]{Albrecht2014CVVel}
{Albrecht}, S., {Winn}, J.~N., {Torres}, G., {et~al.} 2014{\natexlab{a}}, \apj, 785, 83, \dodoi{10.1088/0004-637X/785/2/83}

\bibitem[{{Albrecht} {et~al.}(2014{\natexlab{b}}){Albrecht}, {Winn}, {Torres}, {Fabrycky}, {Setiawan}, {Gillon}, {Jehin}, {Triaud}, {Queloz}, {Snellen}, \& {Eggleton}}]{albrecht2014}
---. 2014{\natexlab{b}}, \apj, 785, 83, \dodoi{10.1088/0004-637X/785/2/83}

\bibitem[{{Albrecht} {et~al.}(2022){Albrecht}, {Dawson}, \& {Winn}}]{AlbrechtDawsonWinn2022}
{Albrecht}, S.~H., {Dawson}, R.~I., \& {Winn}, J.~N. 2022, arXiv e-prints, arXiv:2203.05460.
\newblock \doarXiv{2203.05460}

\bibitem[{{Anderson} \& {Lai}(2021)}]{AndersonLai2021}
{Anderson}, K.~R., \& {Lai}, D. 2021, \apj, 906, 17, \dodoi{10.3847/1538-4357/abcda2}

\bibitem[{{Anderson} {et~al.}(2017){Anderson}, {Lai}, \& {Storch}}]{AndersonLaiStorch2017}
{Anderson}, K.~R., {Lai}, D., \& {Storch}, N.~I. 2017, \mnras, 467, 3066, \dodoi{10.1093/mnras/stx293}

\bibitem[{{Artymowicz} \& {Lubow}(1996)}]{1996arty_circumbinarydisk}
{Artymowicz}, P., \& {Lubow}, S.~H. 1996, \apjl, 467, L77, \dodoi{10.1086/310200}

\bibitem[{{Ball} {et~al.}(2023){Ball}, {Triaud}, {Hatt}, {Nielsen}, \& {Chaplin}}]{Ball+2023}
{Ball}, W.~H., {Triaud}, A. H.~M.~J., {Hatt}, E., {Nielsen}, M.~B., \& {Chaplin}, W.~J. 2023, \mnras, 521, L1, \dodoi{10.1093/mnrasl/slad012}

\bibitem[{{Bashi} {et~al.}(2023){Bashi}, {Mazeh}, \& {Faigler}}]{Bashi+2023}
{Bashi}, D., {Mazeh}, T., \& {Faigler}, S. 2023, \mnras, 522, 1184, \dodoi{10.1093/mnras/stad999}

\bibitem[{{Bate}(2018)}]{bate2018}
{Bate}, M.~R. 2018, \mnras, 475, 5618, \dodoi{10.1093/mnras/sty169}

\bibitem[{{Bate} {et~al.}(2010){Bate}, {Lodato}, \& {Pringle}}]{bate2010}
{Bate}, M.~R., {Lodato}, G., \& {Pringle}, J.~E. 2010, \mnras, 401, 1505, \dodoi{10.1111/j.1365-2966.2009.15773.x}

\bibitem[{{Blanco-Cuaresma} {et~al.}(2014){Blanco-Cuaresma}, {Soubiran}, {Jofr{\'e}}, \& {Heiter}}]{BlancoiSpec2014}
{Blanco-Cuaresma}, S., {Soubiran}, C., {Jofr{\'e}}, P., \& {Heiter}, U. 2014, in Astronomical Society of India Conference Series, Vol.~11, Astronomical Society of India Conference Series, 85--91, \dodoi{10.48550/arXiv.1312.4545}

\bibitem[{{Buder} {et~al.}(2021){Buder}, {Sharma}, {Kos}, {Amarsi}, {Nordlander}, {Lind}, {Martell}, {Asplund}, {Bland-Hawthorn}, {Casey}, {de Silva}, {D'Orazi}, {Freeman}, {Hayden}, {Lewis}, {Lin}, {Schlesinger}, {Simpson}, {Stello}, {Zucker}, {Zwitter}, {Beeson}, {Buck}, {Casagrande}, {Clark}, {{\v{C}}otar}, {da Costa}, {de Grijs}, {Feuillet}, {Horner}, {Kafle}, {Khanna}, {Kobayashi}, {Liu}, {Montet}, {Nandakumar}, {Nataf}, {Ness}, {Spina}, {Tepper-Garc{\'\i}a}, {Ting}, {Traven}, {Vogrin{\v{c}}i{\v{c}}}, {Wittenmyer}, {Wyse}, {{\v{Z}}erjal}, \& {Galah Collaboration}}]{GALAH+}
{Buder}, S., {Sharma}, S., {Kos}, J., {et~al.} 2021, \mnras, 506, 150, \dodoi{10.1093/mnras/stab1242}

\bibitem[{{Djupvik} \& {Andersen}(2010)}]{NOT2010}
{Djupvik}, A.~A., \& {Andersen}, J. 2010, in Astrophysics and Space Science Proceedings, Vol.~14, Highlights of Spanish Astrophysics V, 211, \dodoi{10.1007/978-3-642-11250-8_21}

\bibitem[{{Dorval} {et~al.}(2017){Dorval}, {Boily}, {Moraux}, \& {Roos}}]{2017dorval}
{Dorval}, J., {Boily}, C.~M., {Moraux}, E., \& {Roos}, O. 2017, \mnras, 465, 2198, \dodoi{10.1093/mnras/stw2880}

\bibitem[{{Eggleton} \& {Kiseleva-Eggleton}(2001)}]{eggleton2001}
{Eggleton}, P.~P., \& {Kiseleva-Eggleton}, L. 2001, \apj, 562, 1012, \dodoi{10.1086/323843}

\bibitem[{{Fabrycky} \& {Tremaine}(2007)}]{fabrycky2007}
{Fabrycky}, D., \& {Tremaine}, S. 2007, \apj, 669, 1298, \dodoi{10.1086/521702}

\bibitem[{{Fabrycky} \& {Winn}(2009)}]{fabrycky_winn2009}
{Fabrycky}, D.~C., \& {Winn}, J.~N. 2009, \apj, 696, 1230, \dodoi{10.1088/0004-637X/696/2/1230}

\bibitem[{{Fielding} {et~al.}(2015){Fielding}, {McKee}, {Socrates}, {Cunningham}, \& {Klein}}]{fielding2015}
{Fielding}, D.~B., {McKee}, C.~F., {Socrates}, A., {Cunningham}, A.~J., \& {Klein}, R.~I. 2015, \mnras, 450, 3306, \dodoi{10.1093/mnras/stv836}

\bibitem[{{Fr{\'e}mat} {et~al.}(2023){Fr{\'e}mat}, {Royer}, {Marchal}, {Blomme}, {Sartoretti}, {Guerrier}, {Panuzzo}, {Katz}, {Seabroke}, {Th{\'e}venin}, {Cropper}, {Benson}, {Damerdji}, {Haigron}, {Lobel}, {Smith}, {Baker}, {Chemin}, {David}, {Dolding}, {Gosset}, {Jan{\ss}en}, {Jasniewicz}, {Plum}, {Samaras}, {Snaith}, {Soubiran}, {Vanel}, {Zorec}, {Zwitter}, {Brouillet}, {Caffau}, {Crifo}, {Fabre}, {Fragkoudi}, {Huckle}, {Lasne}, {Leclerc}, {Mastrobuono-Battisti}, {Jean-Antoine Piccolo}, \& {Viala}}]{GaiaVbroad2023}
{Fr{\'e}mat}, Y., {Royer}, F., {Marchal}, O., {et~al.} 2023, \aap, 674, A8, \dodoi{10.1051/0004-6361/202243809}

\bibitem[{{Gaia Collaboration} {et~al.}(2016){Gaia Collaboration}, {Prusti}, {de Bruijne}, {Brown}, {Vallenari}, {Babusiaux}, {Bailer-Jones}, {Bastian}, {Biermann}, {Evans}, {Eyer}, {Jansen}, {Jordi}, {Klioner}, {Lammers}, {Lindegren}, {Luri}, {Mignard}, {Milligan}, {Panem}, {Poinsignon}, {Pourbaix}, {Randich}, {Sarri}, {Sartoretti}, {Siddiqui}, {Soubiran}, {Valette}, {van Leeuwen}, {Walton}, {Aerts}, {Arenou}, {Cropper}, {Drimmel}, {H{\o}g}, {Katz}, {Lattanzi}, {O'Mullane}, {Grebel}, {Holland}, {Huc}, {Passot}, {Bramante}, {Cacciari}, {Casta{\~n}eda}, {Chaoul}, {Cheek}, {De Angeli}, {Fabricius}, {Guerra}, {Hern{\'a}ndez}, {Jean-Antoine-Piccolo}, {Masana}, {Messineo}, {Mowlavi}, {Nienartowicz}, {Ord{\'o}{\~n}ez-Blanco}, {Panuzzo}, {Portell}, {Richards}, {Riello}, {Seabroke}, {Tanga}, {Th{\'e}venin}, {Torra}, {Els}, {Gracia-Abril}, {Comoretto}, {Garcia-Reinaldos}, {Lock}, {Mercier}, {Altmann}, {Andrae}, {Astraatmadja}, {Bellas-Velidis}, {Benson}, {Berthier}, {Blomme}, {Busso}, {Carry}, {Cellino}, {Clementini},
  {Cowell}, {Creevey}, {Cuypers}, {Davidson}, {De Ridder}, {de Torres}, {Delchambre}, {Dell'Oro}, {Ducourant}, {Fr{\'e}mat}, {Garc{\'\i}a-Torres}, {Gosset}, {Halbwachs}, {Hambly}, {Harrison}, {Hauser}, {Hestroffer}, {Hodgkin}, {Huckle}, {Hutton}, {Jasniewicz}, {Jordan}, {Kontizas}, {Korn}, {Lanzafame}, {Manteiga}, {Moitinho}, {Muinonen}, {Osinde}, {Pancino}, {Pauwels}, {Petit}, {Recio-Blanco}, {Robin}, {Sarro}, {Siopis}, {Smith}, {Smith}, {Sozzetti}, {Thuillot}, {van Reeven}, {Viala}, {Abbas}, {Abreu Aramburu}, {Accart}, {Aguado}, {Allan}, {Allasia}, {Altavilla}, {{\'A}lvarez}, {Alves}, {Anderson}, {Andrei}, {Anglada Varela}, {Antiche}, {Antoja}, {Ant{\'o}n}, {Arcay}, {Atzei}, {Ayache}, {Bach}, {Baker}, {Balaguer-N{\'u}{\~n}ez}, {Barache}, {Barata}, {Barbier}, {Barblan}, {Baroni}, {Barrado y Navascu{\'e}s}, {Barros}, {Barstow}, {Becciani}, {Bellazzini}, {Bellei}, {Bello Garc{\'\i}a}, {Belokurov}, {Bendjoya}, {Berihuete}, {Bianchi}, {Bienaym{\'e}}, {Billebaud}, {Blagorodnova}, {Blanco-Cuaresma}, {Boch},
  {Bombrun}, {Borrachero}, {Bouquillon}, {Bourda}, {Bouy}, {Bragaglia}, {Breddels}, {Brouillet}, {Br{\"u}semeister}, {Bucciarelli}, {Budnik}, {Burgess}, {Burgon}, {Burlacu}, {Busonero}, {Buzzi}, {Caffau}, {Cambras}, {Campbell}, {Cancelliere}, {Cantat-Gaudin}, {Carlucci}, {Carrasco}, {Castellani}, {Charlot}, {Charnas}, {Charvet}, {Chassat}, {Chiavassa}, {Clotet}, {Cocozza}, {Collins}, {Collins}, {Costigan}, {Crifo}, {Cross}, {Crosta}, {Crowley}, {Dafonte}, {Damerdji}, {Dapergolas}, {David}, {David}, {De Cat}, {de Felice}, {de Laverny}, {De Luise}, {De March}, {de Martino}, {de Souza}, {Debosscher}, {del Pozo}, {Delbo}, {Delgado}, {Delgado}, {di Marco}, {Di Matteo}, {Diakite}, {Distefano}, {Dolding}, {Dos Anjos}, {Drazinos}, {Dur{\'a}n}, {Dzigan}, {Ecale}, {Edvardsson}, {Enke}, {Erdmann}, {Escolar}, {Espina}, {Evans}, {Eynard Bontemps}, {Fabre}, {Fabrizio}, {Faigler}, {Falc{\~a}o}, {Farr{\`a}s Casas}, {Faye}, {Federici}, {Fedorets}, {Fern{\'a}ndez-Hern{\'a}ndez}, {Fernique}, {Fienga}, {Figueras}, {Filippi},
  {Findeisen}, {Fonti}, {Fouesneau}, {Fraile}, {Fraser}, {Fuchs}, {Furnell}, {Gai}, {Galleti}, {Galluccio}, {Garabato}, {Garc{\'\i}a-Sedano}, {Gar{\'e}}, {Garofalo}, {Garralda}, {Gavras}, {Gerssen}, {Geyer}, {Gilmore}, {Girona}, {Giuffrida}, {Gomes}, {Gonz{\'a}lez-Marcos}, {Gonz{\'a}lez-N{\'u}{\~n}ez}, {Gonz{\'a}lez-Vidal}, {Granvik}, {Guerrier}, {Guillout}, {Guiraud}, {G{\'u}rpide}, {Guti{\'e}rrez-S{\'a}nchez}, {Guy}, {Haigron}, {Hatzidimitriou}, {Haywood}, {Heiter}, {Helmi}, {Hobbs}, {Hofmann}, {Holl}, {Holland}, {Hunt}, {Hypki}, {Icardi}, {Irwin}, {Jevardat de Fombelle}, {Jofr{\'e}}, {Jonker}, {Jorissen}, {Julbe}, {Karampelas}, {Kochoska}, {Kohley}, {Kolenberg}, {Kontizas}, {Koposov}, {Kordopatis}, {Koubsky}, {Kowalczyk}, {Krone-Martins}, {Kudryashova}, {Kull}, {Bachchan}, {Lacoste-Seris}, {Lanza}, {Lavigne}, {Le Poncin-Lafitte}, {Lebreton}, {Lebzelter}, {Leccia}, {Leclerc}, {Lecoeur-Taibi}, {Lemaitre}, {Lenhardt}, {Leroux}, {Liao}, {Licata}, {Lindstr{\o}m}, {Lister}, {Livanou}, {Lobel}, {L{\"o}ffler},
  {L{\'o}pez}, {Lopez-Lozano}, {Lorenz}, {Loureiro}, {MacDonald}, {Magalh{\~a}es Fernandes}, {Managau}, {Mann}, {Mantelet}, {Marchal}, {Marchant}, {Marconi}, {Marie}, {Marinoni}, {Marrese}, {Marschalk{\'o}}, {Marshall}, {Mart{\'\i}n-Fleitas}, {Martino}, {Mary}, {Matijevi{\v{c}}}, {Mazeh}, {McMillan}, {Messina}, {Mestre}, {Michalik}, {Millar}, {Miranda}, {Molina}, {Molinaro}, {Molinaro}, {Moln{\'a}r}, {Moniez}, {Montegriffo}, {Monteiro}, {Mor}, {Mora}, {Morbidelli}, {Morel}, {Morgenthaler}, {Morley}, {Morris}, {Mulone}, {Muraveva}, {Musella}, {Narbonne}, {Nelemans}, {Nicastro}, {Noval}, {Ord{\'e}novic}, {Ordieres-Mer{\'e}}, {Osborne}, {Pagani}, {Pagano}, {Pailler}, {Palacin}, {Palaversa}, {Parsons}, {Paulsen}, {Pecoraro}, {Pedrosa}, {Pentik{\"a}inen}, {Pereira}, {Pichon}, {Piersimoni}, {Pineau}, {Plachy}, {Plum}, {Poujoulet}, {Pr{\v{s}}a}, {Pulone}, {Ragaini}, {Rago}, {Rambaux}, {Ramos-Lerate}, {Ranalli}, {Rauw}, {Read}, {Regibo}, {Renk}, {Reyl{\'e}}, {Ribeiro}, {Rimoldini}, {Ripepi}, {Riva}, {Rixon},
  {Roelens}, {Romero-G{\'o}mez}, {Rowell}, {Royer}, {Rudolph}, {Ruiz-Dern}, {Sadowski}, {Sagrist{\`a} Sell{\'e}s}, {Sahlmann}, {Salgado}, {Salguero}, {Sarasso}, {Savietto}, {Schnorhk}, {Schultheis}, {Sciacca}, {Segol}, {Segovia}, {Segransan}, {Serpell}, {Shih}, {Smareglia}, {Smart}, {Smith}, {Solano}, {Solitro}, {Sordo}, {Soria Nieto}, {Souchay}, {Spagna}, {Spoto}, {Stampa}, {Steele}, {Steidelm{\"u}ller}, {Stephenson}, {Stoev}, {Suess}, {S{\"u}veges}, {Surdej}, {Szabados}, {Szegedi-Elek}, {Tapiador}, {Taris}, {Tauran}, {Taylor}, {Teixeira}, {Terrett}, {Tingley}, {Trager}, {Turon}, {Ulla}, {Utrilla}, {Valentini}, {van Elteren}, {Van Hemelryck}, {van Leeuwen}, {Varadi}, {Vecchiato}, {Veljanoski}, {Via}, {Vicente}, {Vogt}, {Voss}, {Votruba}, {Voutsinas}, {Walmsley}, {Weiler}, {Weingrill}, {Werner}, {Wevers}, {Whitehead}, {Wyrzykowski}, {Yoldas}, {{\v{Z}}erjal}, {Zucker}, {Zurbach}, {Zwitter}, {Alecu}, {Allen}, {Allende Prieto}, {Amorim}, {Anglada-Escud{\'e}}, {Arsenijevic}, {Azaz}, {Balm}, {Beck}, {Bernstein},
  {Bigot}, {Bijaoui}, {Blasco}, {Bonfigli}, {Bono}, {Boudreault}, {Bressan}, {Brown}, {Brunet}, {Bunclark}, {Buonanno}, {Butkevich}, {Carret}, {Carrion}, {Chemin}, {Ch{\'e}reau}, {Corcione}, {Darmigny}, {de Boer}, {de Teodoro}, {de Zeeuw}, {Delle Luche}, {Domingues}, {Dubath}, {Fodor}, {Fr{\'e}zouls}, {Fries}, {Fustes}, {Fyfe}, {Gallardo}, {Gallegos}, {Gardiol}, {Gebran}, {Gomboc}, {G{\'o}mez}, {Grux}, {Gueguen}, {Heyrovsky}, {Hoar}, {Iannicola}, {Isasi Parache}, {Janotto}, {Joliet}, {Jonckheere}, {Keil}, {Kim}, {Klagyivik}, {Klar}, {Knude}, {Kochukhov}, {Kolka}, {Kos}, {Kutka}, {Lainey}, {LeBouquin}, {Liu}, {Loreggia}, {Makarov}, {Marseille}, {Martayan}, {Martinez-Rubi}, {Massart}, {Meynadier}, {Mignot}, {Munari}, {Nguyen}, {Nordlander}, {Ocvirk}, {O'Flaherty}, {Olias Sanz}, {Ortiz}, {Osorio}, {Oszkiewicz}, {Ouzounis}, {Palmer}, {Park}, {Pasquato}, {Peltzer}, {Peralta}, {P{\'e}turaud}, {Pieniluoma}, {Pigozzi}, {Poels}, {Prat}, {Prod'homme}, {Raison}, {Rebordao}, {Risquez}, {Rocca-Volmerange}, {Rosen},
  {Ruiz-Fuertes}, {Russo}, {Sembay}, {Serraller Vizcaino}, {Short}, {Siebert}, {Silva}, {Sinachopoulos}, {Slezak}, {Soffel}, {Sosnowska}, {Strai{\v{z}}ys}, {ter Linden}, {Terrell}, {Theil}, {Tiede}, {Troisi}, {Tsalmantza}, {Tur}, {Vaccari}, {Vachier}, {Valles}, {Van Hamme}, {Veltz}, {Virtanen}, {Wallut}, {Wichmann}, {Wilkinson}, {Ziaeepour}, \& {Zschocke}}]{GaiaMission}
{Gaia Collaboration}, {Prusti}, T., {de Bruijne}, J.~H.~J., {et~al.} 2016, \aap, 595, A1, \dodoi{10.1051/0004-6361/201629272}

\bibitem[{{Gaia Collaboration} {et~al.}(2021){Gaia Collaboration}, {Brown}, {Vallenari}, {Prusti}, {de Bruijne}, {Babusiaux}, {Biermann}, {Creevey}, {Evans}, {Eyer}, {Hutton}, {Jansen}, {Jordi}, {Klioner}, {Lammers}, {Lindegren}, {Luri}, {Mignard}, {Panem}, {Pourbaix}, {Randich}, {Sartoretti}, {Soubiran}, {Walton}, {Arenou}, {Bailer-Jones}, {Bastian}, {Cropper}, {Drimmel}, {Katz}, {Lattanzi}, {van Leeuwen}, {Bakker}, {Cacciari}, {Casta{\~n}eda}, {De Angeli}, {Ducourant}, {Fabricius}, {Fouesneau}, {Fr{\'e}mat}, {Guerra}, {Guerrier}, {Guiraud}, {Jean-Antoine Piccolo}, {Masana}, {Messineo}, {Mowlavi}, {Nicolas}, {Nienartowicz}, {Pailler}, {Panuzzo}, {Riclet}, {Roux}, {Seabroke}, {Sordo}, {Tanga}, {Th{\'e}venin}, {Gracia-Abril}, {Portell}, {Teyssier}, {Altmann}, {Andrae}, {Bellas-Velidis}, {Benson}, {Berthier}, {Blomme}, {Brugaletta}, {Burgess}, {Busso}, {Carry}, {Cellino}, {Cheek}, {Clementini}, {Damerdji}, {Davidson}, {Delchambre}, {Dell'Oro}, {Fern{\'a}ndez-Hern{\'a}ndez}, {Galluccio}, {Garc{\'\i}a-Lario},
  {Garcia-Reinaldos}, {Gonz{\'a}lez-N{\'u}{\~n}ez}, {Gosset}, {Haigron}, {Halbwachs}, {Hambly}, {Harrison}, {Hatzidimitriou}, {Heiter}, {Hern{\'a}ndez}, {Hestroffer}, {Hodgkin}, {Holl}, {Jan{\ss}en}, {Jevardat de Fombelle}, {Jordan}, {Krone-Martins}, {Lanzafame}, {L{\"o}ffler}, {Lorca}, {Manteiga}, {Marchal}, {Marrese}, {Moitinho}, {Mora}, {Muinonen}, {Osborne}, {Pancino}, {Pauwels}, {Petit}, {Recio-Blanco}, {Richards}, {Riello}, {Rimoldini}, {Robin}, {Roegiers}, {Rybizki}, {Sarro}, {Siopis}, {Smith}, {Sozzetti}, {Ulla}, {Utrilla}, {van Leeuwen}, {van Reeven}, {Abbas}, {Abreu Aramburu}, {Accart}, {Aerts}, {Aguado}, {Ajaj}, {Altavilla}, {{\'A}lvarez}, {{\'A}lvarez Cid-Fuentes}, {Alves}, {Anderson}, {Anglada Varela}, {Antoja}, {Audard}, {Baines}, {Baker}, {Balaguer-N{\'u}{\~n}ez}, {Balbinot}, {Balog}, {Barache}, {Barbato}, {Barros}, {Barstow}, {Bartolom{\'e}}, {Bassilana}, {Bauchet}, {Baudesson-Stella}, {Becciani}, {Bellazzini}, {Bernet}, {Bertone}, {Bianchi}, {Blanco-Cuaresma}, {Boch}, {Bombrun}, {Bossini},
  {Bouquillon}, {Bragaglia}, {Bramante}, {Breedt}, {Bressan}, {Brouillet}, {Bucciarelli}, {Burlacu}, {Busonero}, {Butkevich}, {Buzzi}, {Caffau}, {Cancelliere}, {C{\'a}novas}, {Cantat-Gaudin}, {Carballo}, {Carlucci}, {Carnerero}, {Carrasco}, {Casamiquela}, {Castellani}, {Castro-Ginard}, {Castro Sampol}, {Chaoul}, {Charlot}, {Chemin}, {Chiavassa}, {Cioni}, {Comoretto}, {Cooper}, {Cornez}, {Cowell}, {Crifo}, {Crosta}, {Crowley}, {Dafonte}, {Dapergolas}, {David}, {David}, {de Laverny}, {De Luise}, {De March}, {De Ridder}, {de Souza}, {de Teodoro}, {de Torres}, {del Peloso}, {del Pozo}, {Delbo}, {Delgado}, {Delgado}, {Delisle}, {Di Matteo}, {Diakite}, {Diener}, {Distefano}, {Dolding}, {Eappachen}, {Edvardsson}, {Enke}, {Esquej}, {Fabre}, {Fabrizio}, {Faigler}, {Fedorets}, {Fernique}, {Fienga}, {Figueras}, {Fouron}, {Fragkoudi}, {Fraile}, {Franke}, {Gai}, {Garabato}, {Garcia-Gutierrez}, {Garc{\'\i}a-Torres}, {Garofalo}, {Gavras}, {Gerlach}, {Geyer}, {Giacobbe}, {Gilmore}, {Girona}, {Giuffrida}, {Gomel}, {Gomez},
  {Gonzalez-Santamaria}, {Gonz{\'a}lez-Vidal}, {Granvik}, {Guti{\'e}rrez-S{\'a}nchez}, {Guy}, {Hauser}, {Haywood}, {Helmi}, {Hidalgo}, {Hilger}, {H{\l}adczuk}, {Hobbs}, {Holland}, {Huckle}, {Jasniewicz}, {Jonker}, {Juaristi Campillo}, {Julbe}, {Karbevska}, {Kervella}, {Khanna}, {Kochoska}, {Kontizas}, {Kordopatis}, {Korn}, {Kostrzewa-Rutkowska}, {Kruszy{\'n}ska}, {Lambert}, {Lanza}, {Lasne}, {Le Campion}, {Le Fustec}, {Lebreton}, {Lebzelter}, {Leccia}, {Leclerc}, {Lecoeur-Taibi}, {Liao}, {Licata}, {Lindstr{\o}m}, {Lister}, {Livanou}, {Lobel}, {Madrero Pardo}, {Managau}, {Mann}, {Marchant}, {Marconi}, {Marcos Santos}, {Marinoni}, {Marocco}, {Marshall}, {Martin Polo}, {Mart{\'\i}n-Fleitas}, {Masip}, {Massari}, {Mastrobuono-Battisti}, {Mazeh}, {McMillan}, {Messina}, {Michalik}, {Millar}, {Mints}, {Molina}, {Molinaro}, {Moln{\'a}r}, {Montegriffo}, {Mor}, {Morbidelli}, {Morel}, {Morris}, {Mulone}, {Munoz}, {Muraveva}, {Murphy}, {Musella}, {Noval}, {Ord{\'e}novic}, {Orr{\`u}}, {Osinde}, {Pagani}, {Pagano},
  {Palaversa}, {Palicio}, {Panahi}, {Pawlak}, {Pe{\~n}alosa Esteller}, {Penttil{\"a}}, {Piersimoni}, {Pineau}, {Plachy}, {Plum}, {Poggio}, {Poretti}, {Poujoulet}, {Pr{\v{s}}a}, {Pulone}, {Racero}, {Ragaini}, {Rainer}, {Raiteri}, {Rambaux}, {Ramos}, {Ramos-Lerate}, {Re Fiorentin}, {Regibo}, {Reyl{\'e}}, {Ripepi}, {Riva}, {Rixon}, {Robichon}, {Robin}, {Roelens}, {Rohrbasser}, {Romero-G{\'o}mez}, {Rowell}, {Royer}, {Rybicki}, {Sadowski}, {Sagrist{\`a} Sell{\'e}s}, {Sahlmann}, {Salgado}, {Salguero}, {Samaras}, {Sanchez Gimenez}, {Sanna}, {Santove{\~n}a}, {Sarasso}, {Schultheis}, {Sciacca}, {Segol}, {Segovia}, {S{\'e}gransan}, {Semeux}, {Shahaf}, {Siddiqui}, {Siebert}, {Siltala}, {Slezak}, {Smart}, {Solano}, {Solitro}, {Souami}, {Souchay}, {Spagna}, {Spoto}, {Steele}, {Steidelm{\"u}ller}, {Stephenson}, {S{\"u}veges}, {Szabados}, {Szegedi-Elek}, {Taris}, {Tauran}, {Taylor}, {Teixeira}, {Thuillot}, {Tonello}, {Torra}, {Torra}, {Turon}, {Unger}, {Vaillant}, {van Dillen}, {Vanel}, {Vecchiato}, {Viala}, {Vicente},
  {Voutsinas}, {Weiler}, {Wevers}, {Wyrzykowski}, {Yoldas}, {Yvard}, {Zhao}, {Zorec}, {Zucker}, {Zurbach}, \& {Zwitter}}]{GaiaDR3}
{Gaia Collaboration}, {Brown}, A.~G.~A., {Vallenari}, A., {et~al.} 2021, \aap, 649, A1, \dodoi{10.1051/0004-6361/202039657}

\bibitem[{{Gaia Collaboration} {et~al.}(2022){Gaia Collaboration}, {Arenou}, {Babusiaux}, {Barstow}, {Faigler}, {Jorissen}, {Kervella}, {Mazeh}, {Mowlavi}, {Panuzzo}, {Sahlmann}, {Shahaf}, {Sozzetti}, {Bauchet}, {Damerdji}, {Gavras}, {Giacobbe}, {Gosset}, {Halbwachs}, {Holl}, {Lattanzi}, {Leclerc}, {Morel}, {Pourbaix}, {Re Fiorentin}, {Sadowski}, {S{\'e}gransan}, {Siopis}, {Teyssier}, {Zwitter}, {Planquart}, {Brown}, {Vallenari}, {Prusti}, {de Bruijne}, {Biermann}, {Creevey}, {Ducourant}, {Evans}, {Eyer}, {Guerra}, {Hutton}, {Jordi}, {Klioner}, {Lammers}, {Lindegren}, {Luri}, {Mignard}, {Panem}, {Randich}, {Sartoretti}, {Soubiran}, {Tanga}, {Walton}, {Bailer-Jones}, {Bastian}, {Drimmel}, {Jansen}, {Katz}, {van Leeuwen}, {Bakker}, {Cacciari}, {Casta{\~n}eda}, {De Angeli}, {Fabricius}, {Fouesneau}, {Fr{\'e}mat}, {Galluccio}, {Guerrier}, {Heiter}, {Masana}, {Messineo}, {Nicolas}, {Nienartowicz}, {Pailler}, {Riclet}, {Roux}, {Seabroke}, {Sordo}, {Th{\'e}venin}, {Gracia-Abril}, {Portell}, {Altmann}, {Andrae},
  {Audard}, {Bellas-Velidis}, {Benson}, {Berthier}, {Blomme}, {Burgess}, {Busonero}, {Busso}, {C{\'a}novas}, {Carry}, {Cellino}, {Cheek}, {Clementini}, {Davidson}, {de Teodoro}, {Nu{\~n}ez Campos}, {Delchambre}, {Dell'Oro}, {Esquej}, {Fern{\'a}ndez-Hern{\'a}ndez}, {Fraile}, {Garabato}, {Garc{\'\i}a-Lario}, {Haigron}, {Hambly}, {Harrison}, {Hern{\'a}ndez}, {Hestroffer}, {Hodgkin}, {Jan{\ss}en}, {Jevardat de Fombelle}, {Jordan}, {Krone-Martins}, {Lanzafame}, {L{\"o}ffler}, {Marchal}, {Marrese}, {Moitinho}, {Muinonen}, {Osborne}, {Pancino}, {Pauwels}, {Recio-Blanco}, {Reyl{\'e}}, {Riello}, {Rimoldini}, {Roegiers}, {Rybizki}, {Sarro}, {Smith}, {Utrilla}, {van Leeuwen}, {Abbas}, {{\'A}brah{\'a}m}, {Abreu Aramburu}, {Aerts}, {Aguado}, {Ajaj}, {Aldea-Montero}, {Altavilla}, {{\'A}lvarez}, {Alves}, {Anders}, {Anderson}, {Anglada Varela}, {Antoja}, {Baines}, {Baker}, {Balaguer-N{\'u}{\~n}ez}, {Balbinot}, {Balog}, {Barache}, {Barbato}, {Barros}, {Bartolom{\'e}}, {Bassilana}, {Becciani}, {Bellazzini}, {Berihuete},
  {Bernet}, {Bertone}, {Bianchi}, {Binnenfeld}, {Blanco-Cuaresma}, {Blazere}, {Boch}, {Bombrun}, {Bossini}, {Bouquillon}, {Bragaglia}, {Bramante}, {Breedt}, {Bressan}, {Brouillet}, {Brugaletta}, {Bucciarelli}, {Burlacu}, {Butkevich}, {Buzzi}, {Caffau}, {Cancelliere}, {Cantat-Gaudin}, {Carballo}, {Carlucci}, {Carnerero}, {Carrasco}, {Casamiquela}, {Castellani}, {Castro-Ginard}, {Chaoul}, {Charlot}, {Chemin}, {Chiaramida}, {Chiavassa}, {Chornay}, {Comoretto}, {Contursi}, {Cooper}, {Cornez}, {Cowell}, {Crifo}, {Cropper}, {Crosta}, {Crowley}, {Dafonte}, {Dapergolas}, {David}, {de Laverny}, {De Luise}, {De March}, {De Ridder}, {de Souza}, {de Torres}, {del Peloso}, {del Pozo}, {Delbo}, {Delgado}, {Delisle}, {Demouchy}, {Dharmawardena}, {Diakite}, {Diener}, {Distefano}, {Dolding}, {Enke}, {Fabre}, {Fabrizio}, {Fedorets}, {Fernique}, {Figueras}, {Fournier}, {Fouron}, {Fragkoudi}, {Gai}, {Garcia-Gutierrez}, {Garcia-Reinaldos}, {Garc{\'\i}a-Torres}, {Garofalo}, {Gavel}, {Gerlach}, {Geyer}, {Gilmore}, {Girona},
  {Giuffrida}, {Gomel}, {Gomez}, {Gonz{\'a}lez-N{\'u}{\~n}ez}, {Gonz{\'a}lez-Santamar{\'\i}a}, {Gonz{\'a}lez-Vidal}, {Granvik}, {Guillout}, {Guiraud}, {Guti{\'e}rrez-S{\'a}nchez}, {Guy}, {Hatzidimitriou}, {Hauser}, {Haywood}, {Helmer}, {Helmi}, {Sarmiento}, {Hidalgo}, {H{\l}adczuk}, {Hobbs}, {Holland}, {Huckle}, {Jardine}, {Jasniewicz}, {Jean-Antoine Piccolo}, {Jim{\'e}nez-Arranz}, {Juaristi Campillo}, {Julbe}, {Karbevska}, {Khanna}, {Kordopatis}, {Korn}, {K{\'o}sp{\'a}l}, {Kostrzewa-Rutkowska}, {Kruszy{\'n}ska}, {Kun}, {Laizeau}, {Lambert}, {Lanza}, {Lasne}, {Le Campion}, {Lebreton}, {Lebzelter}, {Leccia}, {Lecoeur-Taibi}, {Liao}, {Licata}, {Lindstr{\o}m}, {Lister}, {Livanou}, {Lobel}, {Lorca}, {Loup}, {Madrero Pardo}, {Magdaleno Romeo}, {Managau}, {Mann}, {Manteiga}, {Marchant}, {Marconi}, {Marcos}, {Marcos Santos}, {Mar{\'\i}n Pina}, {Marinoni}, {Marocco}, {Marshall}, {Polo}, {Mart{\'\i}n-Fleitas}, {Marton}, {Mary}, {Masip}, {Massari}, {Mastrobuono-Battisti}, {McMillan}, {Messina}, {Michalik}, {Millar},
  {Mints}, {Molina}, {Molinaro}, {Moln{\'a}r}, {Monari}, {Mongui{\'o}}, {Montegriffo}, {Montero}, {Mor}, {Mora}, {Morbidelli}, {Morris}, {Muraveva}, {Murphy}, {Musella}, {Nagy}, {Noval}, {Oca{\~n}a}, {Ogden}, {Ordenovic}, {Osinde}, {Pagani}, {Pagano}, {Palaversa}, {Palicio}, {Pallas-Quintela}, {Panahi}, {Payne-Wardenaar}, {Pe{\~n}alosa Esteller}, {Penttil{\"a}}, {Pichon}, {Piersimoni}, {Pineau}, {Plachy}, {Plum}, {Poggio}, {Pr{\v{s}}a}, {Pulone}, {Racero}, {Ragaini}, {Rainer}, {Raiteri}, {Ramos}, {Ramos-Lerate}, {Regibo}, {Richards}, {Rios Diaz}, {Ripepi}, {Riva}, {Rix}, {Rixon}, {Robichon}, {Robin}, {Robin}, {Roelens}, {Rogues}, {Rohrbasser}, {Romero-G{\'o}mez}, {Rowell}, {Royer}, {Ruz Mieres}, {Rybicki}, {S{\'a}ez N{\'u}{\~n}ez}, {Sagrist{\`a} Sell{\'e}s}, {Salguero}, {Samaras}, {Sanchez Gimenez}, {Sanna}, {Santove{\~n}a}, {Sarasso}, {Schultheis}, {Sciacca}, {Segol}, {Segovia}, {Semeux}, {Siddiqui}, {Siebert}, {Siltala}, {Silvelo}, {Slezak}, {Slezak}, {Smart}, {Snaith}, {Solano}, {Solitro}, {Souami},
  {Souchay}, {Spagna}, {Spina}, {Spoto}, {Steele}, {Steidelm{\"u}ller}, {Stephenson}, {S{\"u}veges}, {Surdej}, {Szabados}, {Szegedi-Elek}, {Taris}, {Taylor}, {Teixeira}, {Tolomei}, {Tonello}, {Torra}, {Torra}, {Torralba Elipe}, {Trabucchi}, {Tsounis}, {Turon}, {Ulla}, {Unger}, {Vaillant}, {van Dillen}, {van Reeven}, {Vanel}, {Vecchiato}, {Viala}, {Vicente}, {Voutsinas}, {Weiler}, {Wevers}, {Wyrzykowski}, {Yoldas}, {Yvard}, {Zhao}, {Zorec}, \& {Zucker}}]{Gaia2022Teaser}
{Gaia Collaboration}, {Arenou}, F., {Babusiaux}, C., {et~al.} 2022, arXiv e-prints, arXiv:2206.05595, \dodoi{10.48550/arXiv.2206.05595}

\bibitem[{{Glebocki} \& {Stawikowski}(1997)}]{glebocki1997}
{Glebocki}, R., \& {Stawikowski}, A. 1997, \aap, 328, 579

\bibitem[{{Guszejnov} {et~al.}(2017){Guszejnov}, {Hopkins}, \& {Krumholz}}]{2017Guszejnov_turb100au}
{Guszejnov}, D., {Hopkins}, P.~F., \& {Krumholz}, M.~R. 2017, \mnras, 468, 4093, \dodoi{10.1093/mnras/stx725}

\bibitem[{{Halbwachs} {et~al.}(2023){Halbwachs}, {Pourbaix}, {Arenou}, {Galluccio}, {Guillout}, {Bauchet}, {Marchal}, {Sadowski}, \& {Teyssier}}]{Halbwachs+2023}
{Halbwachs}, J.-L., {Pourbaix}, D., {Arenou}, F., {et~al.} 2023, \aap, 674, A9, \dodoi{10.1051/0004-6361/202243969}

\bibitem[{{Hale}(1994)}]{hale1994}
{Hale}, A. 1994, \aj, 107, 306, \dodoi{10.1086/116855}

\bibitem[{{Heggie} \& {Rasio}(1996)}]{1996HeggieRasio}
{Heggie}, D.~C., \& {Rasio}, F.~A. 1996, \mnras, 282, 1064, \dodoi{10.1093/mnras/282.3.1064}

\bibitem[{{Howe} \& {Clarke}(2009)}]{howe2009}
{Howe}, K.~S., \& {Clarke}, C.~J. 2009, \mnras, 392, 448, \dodoi{10.1111/j.1365-2966.2008.14073.x}

\bibitem[{{Hwang}(2023)}]{Hwang2023}
{Hwang}, H.-C. 2023, \mnras, 518, 1750, \dodoi{10.1093/mnras/stac3116}

\bibitem[{{Jennings} \& {Chiang}(2021)}]{JenningsChiang2021}
{Jennings}, R.~M., \& {Chiang}, E. 2021, \mnras, 507, 5187, \dodoi{10.1093/mnras/stab2429}

\bibitem[{{Justesen} \& {Albrecht}(2019)}]{JustesenAlbrecht2019}
{Justesen}, A.~B., \& {Albrecht}, S. 2019, \aap, 625, A59, \dodoi{10.1051/0004-6361/201834368}

\bibitem[{{Justesen} \& {Albrecht}(2020)}]{JustesenAlbrecht2020}
---. 2020, \aap, 642, A212, \dodoi{10.1051/0004-6361/202039138}

\bibitem[{{Justesen} \& {Albrecht}(2021)}]{JustesenAlbrecht2021}
---. 2021, \apj, 912, 123, \dodoi{10.3847/1538-4357/abefcd}

\bibitem[{{Kozai}(1962)}]{kozai1962}
{Kozai}, Y. 1962, \aj, 67, 591, \dodoi{10.1086/108790}

\bibitem[{{Kraft}(1967)}]{Kraft1967}
{Kraft}, R.~P. 1967, \apj, 150, 551, \dodoi{10.1086/149359}

\bibitem[{{Lai} \& {Mu{\~n}oz}(2023)}]{2023lai}
{Lai}, D., \& {Mu{\~n}oz}, D.~J. 2023, \araa, 61, 517, \dodoi{10.1146/annurev-astro-052622-022933}

\bibitem[{{Lamers} {et~al.}(2005){Lamers}, {Gieles}, \& {Portegies Zwart}}]{2005lamers_lifetimes}
{Lamers}, H.~J.~G.~L.~M., {Gieles}, M., \& {Portegies Zwart}, S.~F. 2005, \aap, 429, 173, \dodoi{10.1051/0004-6361:20041476}

\bibitem[{{Lehmann} {et~al.}(2013){Lehmann}, {Southworth}, {Tkachenko}, \& {Pavlovski}}]{lehman2013}
{Lehmann}, H., {Southworth}, J., {Tkachenko}, A., \& {Pavlovski}, K. 2013, \aap, 557, A79, \dodoi{10.1051/0004-6361/201321400}

\bibitem[{{Lidov}(1962)}]{lidov1962}
{Lidov}, M.~L. 1962, \planss, 9, 719, \dodoi{10.1016/0032-0633(62)90129-0}

\bibitem[{{Louden} {et~al.}(2021){Louden}, {Winn}, {Petigura}, {Isaacson}, {Howard}, {Masuda}, {Albrecht}, \& {Kosiarek}}]{Louden+2021}
{Louden}, E.~M., {Winn}, J.~N., {Petigura}, E.~A., {et~al.} 2021, \aj, 161, 68, \dodoi{10.3847/1538-3881/abcebd}

\bibitem[{{Marcussen} \& {Albrecht}(2022)}]{MarcussenAlbrecht2022}
{Marcussen}, M.~L., \& {Albrecht}, S.~H. 2022, \apj, 933, 227, \dodoi{10.3847/1538-4357/ac75c2}

\bibitem[{{Masuda} \& {Winn}(2020)}]{MasudaWinn2020}
{Masuda}, K., \& {Winn}, J.~N. 2020, \aj, 159, 81, \dodoi{10.3847/1538-3881/ab65be}

\bibitem[{{Mazeh} \& {Shaham}(1979)}]{mazeh1979}
{Mazeh}, T., \& {Shaham}, J. 1979, \aap, 77, 145

\bibitem[{{Naoz} \& {Fabrycky}(2014)}]{naoz2014}
{Naoz}, S., \& {Fabrycky}, D.~C. 2014, \apj, 793, 137, \dodoi{10.1088/0004-637X/793/2/137}

\bibitem[{{Offner} {et~al.}(2016){Offner}, {Dunham}, {Lee}, {Arce}, \& {Fielding}}]{Offner+2016}
{Offner}, S. S.~R., {Dunham}, M.~M., {Lee}, K.~I., {Arce}, H.~G., \& {Fielding}, D.~B. 2016, \apjl, 827, L11, \dodoi{10.3847/2041-8205/827/1/L11}

\bibitem[{{Offner} {et~al.}(2023){Offner}, {Moe}, {Kratter}, {Sadavoy}, {Jensen}, \& {Tobin}}]{Offner+2023}
{Offner}, S.~S.~R., {Moe}, M., {Kratter}, K.~M., {et~al.} 2023, in Astronomical Society of the Pacific Conference Series, Vol. 534, Protostars and Planets VII, ed. S.~{Inutsuka}, Y.~{Aikawa}, T.~{Muto}, K.~{Tomida}, \& M.~{Tamura}, 275, \dodoi{10.48550/arXiv.2203.10066}

\bibitem[{{Pavlovski} {et~al.}(2011){Pavlovski}, {Southworth}, \& {Kolbas}}]{pavlovski2011}
{Pavlovski}, K., {Southworth}, J., \& {Kolbas}, V. 2011, \apjl, 734, L29, \dodoi{10.1088/2041-8205/734/2/L29}

\bibitem[{{Philippov} \& {Rafikov}(2013)}]{Philippov+2013}
{Philippov}, A.~A., \& {Rafikov}, R.~R. 2013, \apj, 768, 112, \dodoi{10.1088/0004-637X/768/2/112}

\bibitem[{{Rodet} {et~al.}(2021){Rodet}, {Su}, \& {Lai}}]{2021rodetsulai}
{Rodet}, L., {Su}, Y., \& {Lai}, D. 2021, \apj, 913, 104, \dodoi{10.3847/1538-4357/abf8a7}

\bibitem[{{Schlaufman}(2010)}]{schlaufman2010}
{Schlaufman}, K.~C. 2010, \apj, 719, 602, \dodoi{10.1088/0004-637X/719/1/602}

\bibitem[{{Sybilski} {et~al.}(2018){Sybilski}, {Paw{\l}aszek}, {Sybilska}, {Konacki}, {He{\l}miniak}, {Koz{\l}owski}, \& {Ratajczak}}]{sybilski2018}
{Sybilski}, P., {Paw{\l}aszek}, R.~K., {Sybilska}, A., {et~al.} 2018, \mnras, 478, 1942, \dodoi{10.1093/mnras/sty1135}

\bibitem[{{Thies} {et~al.}(2011){Thies}, {Kroupa}, {Goodwin}, {Stamatellos}, \& {Whitworth}}]{Thies+2011}
{Thies}, I., {Kroupa}, P., {Goodwin}, S.~P., {Stamatellos}, D., \& {Whitworth}, A.~P. 2011, \mnras, 417, 1817, \dodoi{10.1111/j.1365-2966.2011.19390.x}

\bibitem[{{Triaud} {et~al.}(2013){Triaud}, {Hebb}, {Anderson}, {Cargile}, {Collier Cameron}, {Doyle}, {Faedi}, {Gillon}, {Gomez Maqueo Chew}, {Hellier}, {Jehin}, {Maxted}, {Naef}, {Pepe}, {Pollacco}, {Queloz}, {S{\'e}gransan}, {Smalley}, {Stassun}, {Udry}, \& {West}}]{triaud2013}
{Triaud}, A.~H.~M.~J., {Hebb}, L., {Anderson}, D.~R., {et~al.} 2013, \aap, 549, A18, \dodoi{10.1051/0004-6361/201219643}

\bibitem[{von Zeipel(1910)}]{vonzeipel}
von Zeipel, H. 1910, Astronomische Nachrichten, 183, 345

\bibitem[{{Wall} {et~al.}(2019){Wall}, {McMillan}, {Mac Low}, {Klessen}, \& {Portegies Zwart}}]{2019wall_capture}
{Wall}, J.~E., {McMillan}, S. L.~W., {Mac Low}, M.-M., {Klessen}, R.~S., \& {Portegies Zwart}, S. 2019, \apj, 887, 62, \dodoi{10.3847/1538-4357/ab4db1}

\bibitem[{{Weis}(1974)}]{weis1974}
{Weis}, E.~W. 1974, \apj, 190, 331, \dodoi{10.1086/152881}

\bibitem[{{Winn} {et~al.}(2017){Winn}, {Petigura}, {Morton}, {Weiss}, {Dai}, {Schlaufman}, {Howard}, {Isaacson}, {Marcy}, {Justesen}, \& {Albrecht}}]{Winn+2017}
{Winn}, J.~N., {Petigura}, E.~A., {Morton}, T.~D., {et~al.} 2017, \aj, 154, 270, \dodoi{10.3847/1538-3881/aa93e3}

\bibitem[{{Xu} {et~al.}(2023){Xu}, {Hwang}, {Hamilton}, \& {Lai}}]{2023xu_superthermal}
{Xu}, S., {Hwang}, H.-C., {Hamilton}, C., \& {Lai}, D. 2023, \apjl, 949, L28, \dodoi{10.3847/2041-8213/acd6f7}

\bibitem[{{Zhou} \& {Huang}(2013)}]{zhou2013}
{Zhou}, G., \& {Huang}, C.~X. 2013, \apjl, 776, L35, \dodoi{10.1088/2041-8205/776/2/L35}

\end{thebibliography}
\bibliographystyle{aasjournal}

\end{document}